%% file: main.tex
\documentclass[letterpaper,twocolumn,10pt]{article}
\usepackage{usenix}
\usepackage[numbers]{natbib}

\usepackage{booktabs}
\usepackage{enumitem}
\usepackage{hyperref}
\usepackage{bm}
\usepackage{tcolorbox}
\usepackage{amsfonts}
\usepackage{mathtools}
\usepackage{amsmath}
\usepackage{amssymb}
\usepackage{amsthm}
\usepackage{xcolor}
\usepackage{wasysym}
\usepackage{multirow}
\usepackage{algorithm}
\usepackage{algorithmic}
\usepackage{makecell}

\newtheorem{definition}{Definition}

\newcommand{\Full}{\CIRCLE}   %
\newcommand{\Half}{\LEFTcircle}
\newcommand{\Empty}{\Circle}    %

\newenvironment{squishitemize}
{\begin{list}{\textbullet}{%
    \setlength{\itemsep}{0pt}%
    \setlength{\parsep}{0pt}%
    \setlength{\topsep}{0pt}%
    \setlength{\parskip}{0pt} %
    \setlength{\labelwidth}{.5in}%
    \setlength{\labelsep}{0.05in} %
    \setlength{\leftmargin}{.15in} %
    }}
  {\end{list}}

  \newenvironment{squishenumerate}
  {\begin{list}{\arabic{enumi}.}{%
    \usecounter{enumi}%
    \setlength{\itemsep}{0pt}%
    \setlength{\parsep}{0pt}%
    \setlength{\topsep}{0pt}%
    \setlength{\parskip}{0pt}%
    \setlength{\labelwidth}{.5in}%
    \setlength{\labelsep}{0.05in}%
    \setlength{\leftmargin}{.2in}}}
  {\end{list}}

\input{trak_macros}

\title{Challenges in Enabling Private Data Valuation}
\author{Yiwei Fu, Tianhao Wang$^*$, Varun Chandrasekaran \\ University of Illinois Urbana-Champaign, $^*$ University of Virginia}

\begin{document}

\maketitle
\input{sections/0_abstract}
\input{sections/1_introduction} 
\input{sections/2_threat}

\input{sections/3_attribution}

\input{sections/4_takeaways}
\input{sections/5_open}
\input{sections/6_conclusion}

\newpage
\bibliographystyle{plainnat}
\bibliography{refs}

\clearpage
\appendix
\section*{Appendix}
\input{sections/app0}
\input{sections/app1}

\end{document}

%% file: trak_macros.tex
\renewcommand{\hat}[1]{\widehat{#1}}

\usepackage{bm}
\newcommand{\btheta}{\bm{\theta}}
\newcommand{\train}{D}
\newcommand{\valid}{D_{\mathrm{valid}}}

\newcommand{\argmin}{\operatorname{argmin}}

%% file: sections/0_abstract.tex
\begin{abstract}

Data valuation methods quantify how individual training examples contribute to a model’s behavior, and are increasingly used for dataset curation, auditing, and emerging data markets. As these techniques become operational, they raise serious privacy concerns: valuation scores can reveal whether a person’s data was included in training, whether it was unusually influential, or what sensitive patterns exist in proprietary datasets. This motivates the study of privacy-preserving data valuation.
However, privacy is fundamentally in tension with valuation utility under differential privacy (DP). DP requires outputs to be insensitive to any single record, while valuation methods are explicitly designed to measure per-record influence. As a result, naive privatization often destroys the fine-grained distinctions needed to rank or attribute value, particularly in heterogeneous datasets where rare examples exert outsized effects.
In this work, we analyze the feasibility of DP-compatible data valuation. We identify the core algorithmic primitives across common valuation frameworks that induce prohibitive sensitivity, explaining why straightforward DP mechanisms fail. We further derive design principles for more privacy-amenable valuation procedures and empirically characterize how privacy constraints degrade ranking fidelity across representative methods and datasets. Our results clarify the limits of current approaches and provide a foundation for developing valuation methods that remain useful under rigorous privacy guarantees.

\end{abstract}

%% file: sections/1_introduction.tex
\section{Introduction}
\label{sec:intro}

\begin{figure}[ht]
    \centering \includegraphics[width=0.9\linewidth]{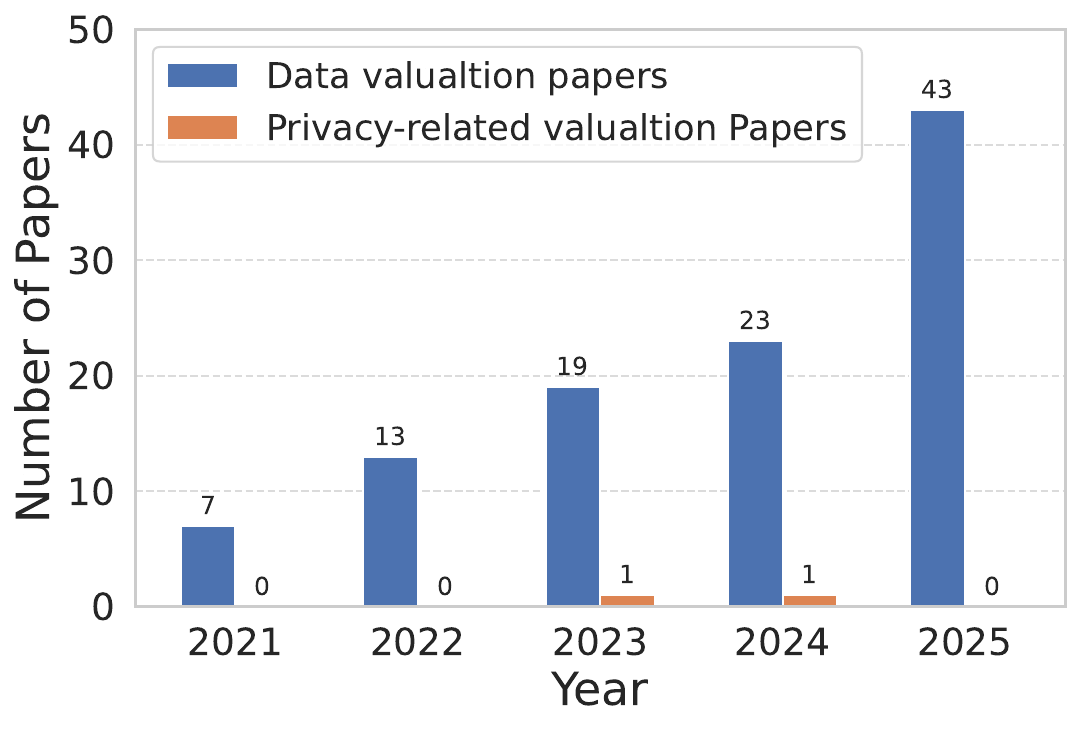}
    \caption{{\bf Private valuation is an understudied topic.} This plot contains the number of accepted data valuation papers, and privacy-focused data valuation papers, over recent years at top ML conferences (NeurIPS, ICLR, ICML). The papers are first filtered with a keyword search, then processed by Gemini to confirm its relevance to data valuation. While the attention on data valuation has grown, little focus was put on privacy-preserving data valuation.}
    \vspace{-0.6cm}
    \label{fig:val_papers}
\end{figure}

Modern machine learning pipelines increasingly rely on \emph{data valuation} to quantify how individual training examples contribute to model performance, robustness, and downstream behavior. As datasets scale and data governance concerns intensify, valuation has become central to dataset curation (e.g., identifying mislabeled or harmful samples)~\cite{Ghorbani_Zou_2019,kwonBetaShapleyUnified}, pricing data in emerging markets~\cite{jia2019efficient,wang2023data}, auditing model failures~\cite{koh2017understanding,basu2021influence}, and attributing responsibility in high-stakes decisions. A wide range of techniques has been proposed, spanning cooperative game–theoretic approaches such as Shapley and Banzhaf values~\cite{Ghorbani_Zou_2019,jia2019efficient,kwonBetaShapleyUnified,wang2023data}, analytic approximations based on influence functions and gradient tracing~\cite{koh2017understanding,Hampel01061974,pruthi2020estimating}, and more recent representation-based surrogates and ``data models’’ that predict model behavior from training subsets~\cite{ilyasDatamodelsPredictingPredictions2022,park2023trak,kim2023gex}.

Despite their methodological diversity, these approaches share a unifying structural objective: estimating how sensitive a model’s behavior is to individual training records. Whether via retraining-based marginal utility, first-order approximations, or learned surrogates, data valuation methods are explicitly designed to detect how adding or removing a single example alters model predictions or loss. This record-level sensitivity is precisely what makes valuation useful for debugging, auditing, and accountability. At the same time, it creates an inherent privacy risk. In many application domains, training data consists of personal, proprietary, or regulated records, and revealing how strongly a particular record influenced a model can itself disclose sensitive information—such as membership, rarity, or identifying attributes.

Differential privacy (DP)~\cite{abadi2016deep} provides a formal framework for protecting against such leakage by requiring that a computation’s output be nearly insensitive to the inclusion or exclusion of any single record. This requirement, however, stands in fundamental tension with the goal of data valuation. Valuation derives meaning from quantifying per-record effects, whereas DP constrains the worst-case magnitude of those effects. As a result, the very signals valuation relies on—per-example gradients, marginal contributions, or influence scores—are precisely those DP mechanisms are designed to suppress. Naively applying DP noise to these quantities often destroys their utility, especially for ranking or identifying the most influential training examples (also the most privacy-sensitive).

This tension raises a basic but unresolved question: \emph{which notions of data valuation, if any, admit meaningful privacy guarantees?}~\autoref{fig:val_papers} illustrates that while data valuation has been extensively studied, comparatively little work directly addresses valuation under formal privacy guarantees. This gap motivates a systematic examination of how existing valuation methods behave when subjected to privacy constraints, and whether their underlying structure is compatible with DP. {\em We aim to better understand that gap}. We focus on the dominant classes of valuation methods\footnote{Often determined based on citations.}, including cooperative game–theoretic approaches (\autoref{sec:marginal}) , influence- (\autoref{sec:if}) and trajectory-based methods (\autoref{sec:traj}), and representation- or surrogate-based techniques (\autoref{sec:linear}) through a unified structural lens. Rather than comparing algorithms solely by accuracy or efficiency, we decompose valuation pipelines into shared primitives such as gradient-based sensitivity, curvature amplification, trajectory accumulation, and subset marginal effects. 

This perspective reveals where prohibitive sensitivity arises, why empirical stabilization techniques (e.g., damping, projection, or averaging) fail to provide dataset-independent guarantees, and why post-hoc privatization is often ineffective. Synthesizing results across the valuation landscape, we distill nine recurring challenges that systematically obstruct DP data valuation (\autoref{sec:lessons}), spanning curvature amplification, coalition extrema, trajectory accumulation, and release-model mismatch. Building on these insights, we further articulate three open problems that delineate promising directions for the community (\autoref{sec:open}): how to decouple individual records from global dataset geometry, how to design valuation objectives with intrinsically bounded sensitivity, and how to formalize release models that align valuation semantics with achievable privacy guarantees. Together, these challenges and open problems clarify which design choices critically determine privacy feasibility and why restricting interaction scope—through locality, bounded neighborhoods, or owner-centric release—can fundamentally alter the privacy–utility tradeoff under DP.

\noindent\textbf{Relation to prior work.}
Existing surveys of data valuation emphasize fairness, efficiency, or economic interpretation~\cite{Pei2022ASO,Sim2022DataValuation,Deng_Hu_Hu_Li_Liu_Wang_Ley_Dai_Huang_Huang_et_al.}, but largely sidestep privacy considerations. While a small number of works study privacy for specific valuation algorithms (e.g., $k$-NN Shapley~\cite{jia2019efficient,wang2023privacy}), ours is the first SoK to systematically analyze the fundamental privacy limitations and design tradeoffs across the full landscape of modern data valuation methods.

%% file: sections/2_threat.tex
\section{Background}

This section introduces the technical and conceptual foundations needed to
reason about data valuation and its interaction with privacy guarantees. We then formalize the threat model and
privacy notions relevant to valuation, clarifying what is protected, what is
released.

\input{sections/2.1_primer}

\input{sections/2.2_tm}

%% file: sections/2.1_primer.tex
\subsection{Machine Learning Primer}

\noindent\textbf{Supervised learning and model training.}
We consider a supervised learning setting in which a model with parameters
$\theta \in \mathbb{R}^d$ maps an input $x$ to a prediction $f(x,\theta)$.
Training is performed on a dataset
$\train = \{z_j = (x_j,y_j)\}_{j=1}^n$
by minimizing an empirical risk objective
\[ 
L(\theta;\train)
=
\frac{1}{n}
\sum_{j=1}^n \ell\!\left(f(x_j,\theta),\,y_j\right),
\]
where $\ell(\cdot,\cdot)$ is a task-specific loss function. In practice, the
optimization is carried out using stochastic gradient descent (SGD) or a
variant thereof. Let $\theta_0$ denote an initialization, and let
$B_t \subseteq \train$ be the mini-batch sampled at iteration $t$. A generic
update takes the form
\[ 
\theta_{t+1}
=
\theta_t
-
\eta_t\,
\nabla_\theta
\Bigg(
\frac{1}{|B_t|}
\sum_{z_j \in B_t}
\ell\!\left(f(x_j,\theta_t),y_j\right)
\Bigg),
\]
yielding a trajectory of parameters $\{\theta_t\}_{t=0}^{T}$ produced by the
training algorithm $\mathcal{A}$. We denote by
\[ 
\btheta
\;\triangleq\;
\theta_T
=
\mathcal{A}(\train)
\]
the final trained parameters, and by $f(x,\btheta)$ the prediction of the
trained model on an input $x$. Many valuation methods studied in this paper
reason about how individual training examples affect either (i) the evolution
of this trajectory, or (ii) the properties of the final model $\btheta$ and its
predictions.

skip

\noindent\textbf{What is data valuation?}
Given a validation example $z=(x,y)\in\valid$, a data valuation method assigns
to each training point $z_j\in\train$ a score $v_j(z)$ that quantifies the
contribution of $z_j$ to the model’s behavior at $z$. Depending on the method,
$v_j(z)$ may capture the change in validation loss under data removal, the
expected marginal utility of $z_j$ across subsets, or the influence of $z_j$
along the optimization trajectory. Later sections instantiate this general
notion through concrete estimators (influence-based, Shapley-style,
trajectory-based, and surrogate methods).

%% file: sections/2.2_tm.tex
\subsection{Threat Model}

We study a threat model centered on the computation and release of data
valuation scores, with emphasis on the privacy risks that arise when such
mechanisms are deployed.

\smallskip
\noindent{\bf Why should valuation be private?}
Valuation scores can encode highly sensitive information about individual
training records. Large or distinctive values may reveal that a record was
included in training, that it is rare or atypical, or that it exerted outsized
influence on model behavior. In regulated domains such as healthcare, finance,
education, or user-generated content, such leakage may violate legal,
contractual, or ethical privacy requirements~\cite{hipaa, gdpr}. In data market settings,
valuation outputs can also expose sensitive economic information about dataset
composition or value distributions. Even in security and auditing workflows,
scores may inadvertently reveal internal data artifacts or vulnerabilities.
Moreover, valuation is often used in interactive or iterative regimes, where
repeated queries, comparisons across models, or aggregation over validation
points can amplify leakage over time. Even approximate or noisy outputs may be
combined to reconstruct sensitive information. Protecting valuation mechanisms
is therefore necessary not only to safeguard individual contributors, but also
to preserve the confidentiality of datasets and deployed models.

\smallskip
\begin{definition}[\textbf{Differential Privacy~\cite{Dwork_Roth_2014}}]
A randomized mechanism $\mathcal{M}$ with domain $\mathcal{Z}^n$ is
$(\varepsilon,\delta)$-\emph{differentially private} if for all pairs of
\emph{neighboring datasets} $\train,\train' \in \mathcal{Z}^n$ that differ in
exactly one training record, and for all measurable output events
$\mathcal{O}$,
\[
\Pr\!\left[\mathcal{M}(\train) \in \mathcal{O}\right]
\;\le\;
e^{\varepsilon}\Pr\!\left[\mathcal{M}(\train') \in \mathcal{O}\right]
\;+\;
\delta .
\]
\end{definition}

This definition formalizes a worst-case notion of stability: the distribution
of the mechanism’s output should not change significantly when any single
training example is modified or removed. Differential privacy (DP) is preserved
under post-processing and composes across multiple adaptive queries, making it
the standard framework for reasoning about privacy in iterative and interactive
learning systems.

\smallskip
\noindent{\bf Two privacy notions for valuation.} A recurring source of confusion in
\emph{private data valuation} is that different works implicitly adopt
different privacy notions and release models, depending on what is protected
and who observes the outputs.
We distinguish two canonical settings.

\begin{squishitemize}
    \item \underline{\em Central release (dataset-level outputs).}
    A mechanism releases a score vector (or ranking) derived from the private
    training set $\train=\{z_1,\dots,z_n\}$ to an analyst. The natural goal is
    standard record-level $(\epsilon,\delta)$-DP with respect to changes in
    \emph{any} single training example.

    \item \underline{\em Per-record release (query-centric / joint-style).}
    A mechanism returns the value of a particular record $z_j$ \emph{to the
    owner of $z_j$}. In this setting it is natural to treat the queried record
    as public to the recipient, while still protecting the remainder of the
    dataset $\train\setminus\{z_j\}$. This one-sided/joint viewpoint
    allows the output for index $j$ to depend arbitrarily on $z_j$, but
    requires that it not reveal much about any other $z_i$. Guarantees proved
    in this per-record setting generally \emph{do not} imply that one can
    publish the entire score vector under standard central DP. {\em Most of the discussion (and methods) in this paper is assumed to be in this setting.}
\end{squishitemize}

%% file: sections/3_attribution.tex
\section{Methods of Data Valuation}

Leave-One-Out (LOO) retraining is the most direct and conceptually faithful
approach to data valuation. Given a training set $\train=\{z_1,\dots,z_n\}$ and
learning algorithm $\mathcal{A}$, let $\btheta=\mathcal{A}(\train)$. For a
candidate point $z_j$, we retrain on $\train\setminus\{z_j\}$ to obtain
$\btheta'=\mathcal{A}(\train\setminus\{z_j\})$, and define its value through
the induced change in an evaluation functional such as loss:
\[
v_j(z)=\ell(z,\btheta)-\ell(z,\btheta').
\]
The quantity $v_j$ measures the marginal contribution of $z_j$ to the learned
model, capturing how much the model’s behavior depends on that record.
LOO valuation lies in the neighboring-dataset perturbation regime central to
algorithmic stability and differential privacy. In particular, $v_j$
constitutes a form of \emph{local sensitivity}: it quantifies the realized,
data-dependent effect of removing a single training example along a chosen
utility.

As discussed in our threat model, releasing such dependence scores can itself
be privacy-sensitive, motivating the need for privacy-preserving estimators.
This leads to our central question: \emph{how can we approximate LOO-style
valuation functions at scale while preserving rigorous privacy guarantees?} We
study influence methods~\cite{koh2017understanding,  kwon2023datainf, martens2015optimizing, schioppa2021scalinginfluencefunctions}, marginal-contribution estimators~\cite{Ghorbani_Zou_2019, kwonBetaShapleyUnified, wang2023data}, trajectory-based attribution~\cite{pruthi2020estimatingtrainingdatainfluence, bae2024trainingdataattributionapproximate, wang2024data}, and surrogate models~\cite{ilyasDatamodelsPredictingPredictions2022, park2023trak} as computational approximations to LOO sensitivity, beginning with influence-based approaches that model deletion effects through first-order parameter perturbations

\input{tables/master}

\input{sections/3.1_influence}
\input{sections/3.2_shap}
\input{sections/3.3_traj}
\input{sections/3.4_surrogate}

%% file: tables/master.tex
\begin{table*}[t]
\centering
    \resizebox{0.99\textwidth}{!}{
\begin{tabular}{l
c c c c c
c c c}
\toprule
\textbf{Method} &

\textbf{Retrain} &
\textbf{Exposure} &
\textbf{Util-Stable} &
\textbf{Sens-Driver} &
\textbf{Scale} &

\multicolumn{3}{c}{\textbf{Privacy--Sensitivity Characteristics}}
\\
\cmidrule(lr){7-9}
 &  &  &  &  &
 & \textbf{Emp-Smooth} & \textbf{Formal-Bound} & \textbf{Clip-Amenable}
\\
\midrule

\multicolumn{9}{l}{\textbf{First-Order Influence and Curvature Surrogates}} \\
\midrule

Exact influence / iHVP
& \Empty & \Empty & \Half & \Full & \Empty
& \Empty & \Empty & \Half
\\

{\bf A1.} Iterative HVP (CG, LiSSA)
& \Empty & \Half & \Half & \Full & \Half
& \Half & \Empty & \Half
\\

{\bf A2.} Fisher-based substitutes ((E)K-FAC, DataInf)
& \Empty & \Empty & \Half & \Half & \Full
& \Full & \Empty & \Half
\\

{\bf A3.} Projection / subspace influence
& \Empty & \Empty & \Half & \Half & \Half
& \Full & \Empty & \Half
\\

\midrule
\multicolumn{9}{l}{\textbf{Marginal-Contribution / Coalition-Based}} \\
\midrule

{\bf A4.} Shapley (Data Shapley)
& \Full & \Full & \Empty & \Full & \Empty
& \Half & \Empty & \Half
\\

{\bf A5.} Beta Shapley
& \Full & \Full & \Empty & \Full & \Empty
& \Half & \Empty & \Half
\\

{\bf A6.} Data Banzhaf
& \Full & \Full & \Empty & \Full & \Empty
& \Half & \Empty & \Half
\\

\midrule

\multicolumn{9}{l}{\textbf{Trajectory-Based Valuation}} \\
\midrule

{\bf A7.} TracIn
& \Empty & \Full & \Half & \Half & \Full
& \Full & \Empty & \Half
\\

{\bf A8.} SOURCE (segmented propagation)
& \Empty & \Full & \Half & \Full & \Half
& \Full & \Empty & \Half
\\

{\bf A9.} In-run Data Shapley
& \Empty & \Full & \Half & \Half & \Half
& \Half & \Empty & \Half
\\
\midrule

\multicolumn{9}{l}{\textbf{Surrogates and Linearization}} \\
\midrule

{\bf A10.} TRAK (Taylor linearization)
& \Empty & \Empty & \Half & \Half & \Full
& \Full & \Empty & \Half
\\

\midrule
\multicolumn{9}{l}{\textbf{Sensitivity-Constrained / Locality-Based}} \\
\midrule

{\bf A11.} T\texttt{-}$k$NN (truncated neighborhood valuation)
& \Empty & \Half & \Full & \Half & \Full
& \Half & \Full & \Full
\\

\bottomrule
\end{tabular}
}
\caption{
\textbf{Privacy-oriented systematization of data valuation approaches.}
Each row corresponds to a valuation family and each column highlights
structural properties that determine sensitivity and DP feasibility. Symbols
denote qualitative support:
\Full\;=\;strongly present or explicitly enforced,\;
\Half\;=\;partially present, heuristic, or context-dependent,\;
\Empty\;=\;absent.
\textbf{Retrain} indicates whether valuation requires repeated retraining.
\textbf{Exposure} captures whether the method involves repeated access across
training steps or coalition evaluations (important for DP composition).
\textbf{Util-Stable} reflects whether the underlying utility admits bounded or
Lipschitz sensitivity.
\textbf{Sens-Driver} identifies the dominant amplification mechanism (e.g.,
Hessian inversion, coalition extrema, or trajectory accumulation).
\textbf{Scale} summarizes feasibility at modern model sizes.
The \textbf{Privacy--Sensitivity} columns characterize privacy behavior:
\emph{Emp-Smooth} denotes empirical attenuation via averaging, projection, or
surrogates;
\emph{Formal-Bound} indicates existence of a dataset-independent worst-case
sensitivity bound;
\emph{Clip-Amenable} captures whether clipping can enforce bounded influence
without destroying valuation semantics. More details are in~\autoref{app:table_explanation}.
\vspace{-5mm}}
\label{tab:sok_unified_privacy_block}
\end{table*}

%% file: sections/3.1_influence.tex
\subsection{Influence \& Curvature Approximations}
\label{sec:if}

Influence functions are a classical tool from robust statistics~\cite{Hampel01061974, Cook_Weisberg_1982} for quantifying how an empirical risk
minimizer changes under infinitesimal perturbations to the training
distribution. In our setting, they provide a scalable \emph{first-order
surrogate} for leave-one-out (LOO) valuation by approximating the effect of
removing a single training example. Let
\[ 
\btheta = \argmin_{\theta}\,\frac{1}{n}\sum_{i=1}^n \ell(z_i,\theta)
\]
denote the ERM solution. Upweighting $z_j$ by $\varepsilon$ yields
\[ 
\hat{\btheta}(\varepsilon)
=
\argmin_{\theta}
\left(
\frac{1}{n}\sum_{i=1}^n \ell(z_i,\theta)
+
\varepsilon\,\ell(z_j,\theta)
\right),
\]
and influence approximates the resulting parameter shift via a first-order
expansion around $\btheta$. Modeling deletion by $\varepsilon\approx -1/n$
yields a local linear approximation to retraining
\cite{koh2017understanding}.

A useful unified view is that influence always takes the form
\begin{equation}\label{eq:if_unified}
v_j(z)
=
-\nabla_\theta \ell(z,\btheta)^\top \;
\underbrace{\mathcal{A}_\lambda}_{\text{inverse-curvature operator}}
\;
\nabla_\theta \ell(z_j,\btheta),
\end{equation}
where $\mathcal{A}_\lambda$ propagates the training-point gradient through an
approximate inverse curvature operator.

In the classical setting,
$\mathcal{A}_\lambda = H_\lambda^{-1}$ with
\[ 
H
=
\frac{1}{n}\sum_{i=1}^n \nabla^2_{\theta}\ell(z_i,\btheta),
\qquad
H_\lambda = H + \lambda I.
\]
Damping ($\lambda>0$) is essential in deep networks due to ill-conditioning and
non-convexity. Computationally, the bottleneck is applying
$H_\lambda^{-1}$ to $\nabla_\theta \ell(z_j,\btheta)$.
Crucially, Eq.~\eqref{eq:if_unified} also isolates the privacy challenge.
In DP valuation, the relevant global sensitivity is governed by
\[ 
\sup_{z_j}\big\|\mathcal{A}_\lambda \nabla_\theta \ell(z_j,\btheta)\big\|,
\]
i.e., how much a single example can be amplified through the chosen inverse
operator. Existing scalable estimators can be interpreted as substituting
different approximations for $\mathcal{A}_\lambda$; while these often smooth
influence values in practice, they do not provide worst-case sensitivity
control.

\smallskip

\noindent \textbf{Approach 1: Iterative estimation via HVPs.}
Iterative methods approximate $\mathcal{A}_\lambda$ by implicitly applying
$H_\lambda^{-1}$ through a truncated solve. Conjugate Gradient (CG)~\cite{martens2010deep} and LiSSA~\cite{agarwal2017second} estimate
\[ 
u_j \approx H_\lambda^{-1}\nabla_\theta \ell(z_j,\btheta)
\]
by solving
\[ 
H_\lambda u_j = \nabla_\theta \ell(z_j,\btheta)
\]
using only Hessian-vector products. In Eq.~\eqref{eq:if_unified}, this
corresponds to an operator
$\mathcal{A}_\lambda \approx \mathsf{Solve}_T(H_\lambda,\cdot)$ implemented by
$T$ solver iterations.

\smallskip

\noindent \textbf{Approach 2: Fisher-based substitute estimators.}
A second family replaces the empirical Hessian with a Fisher surrogate. For
log-likelihood losses, Bartlett's identity motivates approximating curvature by
the Fisher information
\[ 
F
=
\mathbb{E}\big[
\nabla_\theta \ell(z,\btheta)\,
\nabla_\theta^\top \ell(z,\btheta)
\big].
\]
In the unified form, Fisher methods substitute
\[ 
\mathcal{A}_\lambda = (F_\lambda)^{-1},
\qquad F_\lambda = F + \lambda I,
\]
in place of $H_\lambda^{-1}$.
Since $F$ is positive semidefinite and reflects smoother population-level
geometry, Fisher-based operators often attenuate sharp empirical curvature and
stabilize influence estimates. Methods such as (E)K-FAC and \textsc{DataInf}~\cite{martens2015optimizing, grosse2023studyinglargelanguagemodel, kwon2023datainf}
exploit layerwise structure to approximate $(F_\lambda)^{-1}$ efficiently.

\smallskip

\noindent \textbf{Approach 3: Subspace projections.}
Projection methods approximate inverse curvature by restricting computation to
a low-dimensional subspace. Given an orthonormal basis
$G\in\mathbb{R}^{d\times m}$, they use the projected operator
\[ 
\mathcal{A}_\lambda
=
G\,(G^\top H_\lambda G)^{-1}G^\top,
\]
so Eq.~\eqref{eq:if_unified} propagates gradients only along the spectral
directions captured by the subspace. Krylov/Arnoldi constructions
\cite{schioppa2021scalinginfluencefunctions} provide scalable bases using only
HVP access.
Projection introduces structured truncation: discarded directions cannot
contribute to influence, which often improves numerical stability. 

\smallskip

\noindent
\textbf{Summary.}
Eq.~\eqref{eq:if_unified} shows that scalable influence estimators differ
primarily in their choice of inverse-curvature operator $\mathcal{A}_\lambda$.
Iterative, Fisher, and projection-based approximations all introduce implicit
regularization that can stabilize influence empirically, but none directly
controls the worst-case amplification of a single example required for
differentially private valuation. In particular, DP mechanisms require a
certified bound on the global sensitivity
$\sup_{z_j}\|\mathcal{A}_\lambda \nabla_\theta \ell(z_j,\btheta)\|$, whereas
these operators are shaped by data-dependent curvature, truncation, or
subspace selection. As a result, rare examples or sharp directions of the loss
geometry can still induce disproportionately large valuation shifts, and
privacy noise calibrated downstream may be insufficient unless sensitivity is
explicitly bounded. This highlights a fundamental gap: scalable influence
approximations provide computational tractability and practical smoothing, but
additional mechanisms are needed to ensure worst-case stability and formal DP
guarantees.

\paragraph{Privacy driver (curvature amplification).}
In~\autoref{tab:sok_unified_privacy_block}, influence-based methods are
dominated by a \emph{Hessian amplification} sensitivity driver: the worst-case
effect of a single example is governed by
$\sup_{z_j}\|\mathcal{A}_\lambda \nabla_\theta \ell(z_j,\btheta)\|$, which can
become large in ill-conditioned directions even when empirical smoothing is
observed.

%% file: sections/3.2_shap.tex
\subsection{Weighted Marginal Contributions}
\label{sec:marginal}

Rather than modeling how
a point alters the optimization path, these approaches define value as the
\emph{expected marginal contribution} of $z_j$ to a utility function $U$
(e.g., validation accuracy or loss) when added to subsets of the training set.
Concretely, these methods aggregate marginal utilities of the form
\[ 
\Delta_U(z_j;S)
\;=\;
U(S\cup\{z_j\}) - U(S),
\]
averaged over different subset-selection rules. Because exact enumeration over
all coalitions is infeasible, practical estimators differ in how they sample
and weight subsets, reflecting distinct statistical and robustness tradeoffs.

This formulation also makes the privacy challenge explicit. Differential
privacy depends on controlling the \emph{worst-case} marginal effect of a
single example,
\[ 
\sup_{S,z_j}\big|\Delta_U(z_j;S)\big|,
\]
whereas Shapley-style estimators are defined as expectations over subsets. As a
result, even when average marginal contributions are small, rare coalitions
can induce large utility jumps, yielding high global sensitivity and
complicating DP noise calibration. We discuss three representative families
below.

\smallskip

\noindent \textbf{Approach 4: Classical Shapley-style data valuation.}
\cite{Ghorbani_Zou_2019} adapt the Shapley value from cooperative game theory by
defining the value of a point $z_j$ as its average marginal contribution to a
utility function $U$ across all subsets:
{
\begin{equation}\label{eq:data_shapley}
v_j
=
\sum_{k=1}^{n}
\frac{1}{n}
\binom{n-1}{k-1}^{-1}
\!\!\!\!
\sum_{\substack{S \subseteq \train \setminus \{z_j\},\\ |S|=k-1}}
\big(
U(S \cup \{z_j\}) - U(S)
\big).
\end{equation}
}
This definition is axiomatically appealing (symmetry, efficiency, and dummy),
and provides a clear semantic notion of value as expected marginal utility.
Exact computation requires evaluating $U$ over exponentially many coalitions,
so practical implementations rely on stochastic estimators, most commonly
permutation sampling. A random ordering of the training set is drawn, and the
marginal contribution of $z_j$ is computed relative to the prefix preceding it.
Averaging across permutations yields an unbiased estimator, but with variance
driven by both subset randomness and stochasticity in training and evaluation.

The formulation Eq.~\eqref{eq:data_shapley} mandates the iteration over all possible subsets of $D \setminus \{z_j\}$, which scales exponentially with dataset size and is practically infeasible to compute for any dataset with a reasonable size. Thus multiple approximation is commonly used in practice.

\begin{squishitemize}
    \item Naive Monte Carlo (MC) Shapley is a direct Monte Carlo estimation of data Shapley value by sampling random subsets $S \subseteq D \setminus \{z_j\}$. However, this method usually has poor coverage.
    \item Stratified Shapley~\cite{Wu_Jia_Lin_Huang_Chang_2023} first samples a size $k \in \{0, 1, \ldots, n-1\}$, then sample subsets of given size $k$. This ensure a controlled representation of dataset of different sizes. The sampling distribution of $k$ can be uniform, or customized.
    \item Permutation-based Shapley formulates the Shapley value by modeling training randomness through data orderings. The contribution of a data point is measured as its marginal effect on the utility when it is added to a prefix of a permutation of the dataset i.e.
    \[
    v_j
    =
    \frac{1}{n!}
    \sum_{\sigma \in \Pi(n)}
    \big(
    U(S^\sigma_j \cup \{z_j\}) - U(S^\sigma_j)
    \big)
    \]
    where $\Pi(S)$ denotes all permutations of $S$, and $S^\sigma_j$ denotes the set of indices in permutation $\sigma$ before the position where $z_j$ appears. Even though the formulation involves $n!$ terms, the sample complexity is polynomial~\cite{Castro_Gomez_Tejada_2009}.
    \item Truncated MC~\cite{Ghorbani_Zou_2019} is actually a permutation based Shapley which further speeds up the calculation. By adding early stopping in calculating $\Delta_U(z_j;S^\sigma)$ for a given permutation $\sigma$, this method avoids retraining on the full dataset for every sampled permutation $\sigma$.
\end{squishitemize}

\smallskip

\noindent \textbf{Approach 5: Beta Shapley (size-weighted marginal contributions).}
Kwon et al.~\cite{kwonBetaShapleyUnified} generalize the Shapley formulation by
observing that its weighting over subset sizes is only one of many valid
aggregation rules. They define Beta Shapley by assigning deterministic weights
$w(k)$ to coalitions of size $k-1$:
\[ 
v_j
=
\frac{1}{n}
\sum_{k=1}^{n}
w(k)
\!\!\!\!
\sum_{\substack{S \subseteq \train \setminus \{z_j\},\\ |S|=k-1}}
\big(
U(S \cup \{z_j\}) - U(S)
\big),
\]
with normalization ensuring consistency. Choosing
\[ 
w_{\alpha,\beta}(k)
=
n\,\frac{B(k+\beta-1,\,n-k+\alpha)}{B(\alpha,\beta)}
\]
yields a two-parameter family that interpolates between emphasizing early- and
late-joining contributions.
This flexibility allows practitioners to prioritize particular coalition
regimes (e.g., scarce-data settings), but it also reshapes the sensitivity
profile of the estimator. Reweighting toward specific subset sizes can amplify
marginal contributions precisely in those regimes, making the worst-case
sensitivity dependent on both the chosen weighting function and the data. 

\smallskip

\noindent \textbf{Approach 6: Data Banzhaf (uniform subset weighting).}
\cite{wang2023data} argue that in practice the utility function $U$ is often
noisy or stochastic, which can make Shapley-based rankings unstable. They
introduce a safety-margin criterion—requiring rankings to remain unchanged
under small perturbations to $U$—and show that this leads to uniform weighting
over all coalitions. The resulting Data Banzhaf value is
\[ 
v_j
=
\frac{1}{2^{\,n-1}}
\sum_{S \subseteq \train \setminus \{z_j\}}
\big(
U(S \cup \{z_j\}) - U(S)
\big).
\]

Uniform weighting can stabilize empirical rankings by averaging over a broad
range of subsets. However, from a privacy perspective, it also increases
exposure to extreme coalitions. Because no subsets are down-weighted, the
estimator directly incorporates regimes in which the marginal effect of a
single example is unusually large. 

\smallskip

\noindent
\textbf{Summary.}
Marginal contribution methods provide a clear semantic notion of data value as
expected utility improvement across coalitions. However, differential privacy
depends on bounding the worst-case marginal effect of a single example across
all subsets. Classical Shapley, Beta Shapley, and Data Banzhaf estimators differ
in how they weight coalitions, but all define value through averages that can
mask rare yet extreme marginal contributions, leaving global sensitivity
uncontrolled without additional bounding mechanisms.

\paragraph{Privacy driver (utility instability and coalition extrema).}
Coalition-based methods inherit sensitivity directly from the utility
functional: DP depends on bounding
$\sup_{S,z_j}|U(S\cup\{z_j\})-U(S)|$.
Thus, even when valuations average marginal contributions, unstable or
unbounded utilities (e.g., stochastic validation loss) allow rare coalitions to
dominate worst-case sensitivity, as reflected by the ``Util-Stable'' and
``Sens-Driver'' columns in Table~\ref{tab:sok_unified_privacy_block}.

%% file: sections/3.3_traj.tex
\subsection{Trajectory-Aware Approximations}
\label{sec:traj}

Modern deep networks are obtained through long, stochastic
optimization trajectories, during which examples may exert influence in
different ways at different stages of training. This motivates
\emph{trajectory-based} valuation methods that treat the optimization path
itself as the channel through which data causally shapes the final model.
Rather than asking how parameters would differ if $z_j$ were removed at the
end of training, these approaches attribute contribution across the sequence
of intermediate iterates $\{\theta_t\}$ produced by the training algorithm
$\mathcal{A}$.
A useful unified view is that trajectory-based valuations can often be written
as a sum of per-step ``credits'' assigned along the training path:
\begin{multline}\label{eq:traj_unified}
v_j(z)
=
\sum_{t=0}^{T-1}
\eta_t\,\alpha_{t,j}\,
\nabla_\theta \ell(z,\theta_t)^\top
\underbrace{\mathcal{B}^{(1)}_t}_{\text{trajectory propagation}}
\nabla_\theta \ell(z_j,\theta_t) \\
+ \frac{\eta^2}{2}\alpha_{t,j}\nabla_\theta \ell(z_j,\theta_t)^\top
\underbrace{\mathcal{B}^{(2)}_t[z, D]}_{\text{curvature coupling}},
\end{multline}
where $\alpha_{t,j}$ captures when and how strongly $z_j$ participates in
training (e.g., mini-batch exposure or Shapley-style weights), and
$\mathcal{B}^{(1)}_t$ encodes how an update at time $t$ propagates forward through
training dynamics (ranging from the identity operator to segment-level
inverse-curvature and propagation matrices). The second order term $\mathcal{B}^{(2)}_t$ couples training set $D$'s influence at time $t$ with validation sample $z$, and then applies it to the valuation sample $z_j$, but is \textit{usually omitted for simplicity and efficiency}.

Eq.~\eqref{eq:traj_unified} also isolates the privacy difficulty. DP requires controlling the worst-case accumulated contribution of a
single example across the trajectory, which depends on both the exposure
weights $\{\alpha_{t,j}\}$ and the amplification induced by
$\{\mathcal{B}^{(1)}_t, \mathcal{B}^{(2)}_t\}$. Existing trajectory-aware estimators provide scalable attribution, but their per-step credits remain strongly data- and
trajectory-dependent, leaving global sensitivity difficult to bound.

\smallskip

\noindent \textbf{Approach 7: Trajectory gradient similarity (TracIn).}
A first trajectory-aware family departs from Hessian-based influence and
instead attributes contribution through gradient alignment along the training
path. The key intuition in Pruthi et al.~\cite{pruthi2020estimating} is that under SGD, the effect of a point $z_j$ is mediated through the update directions it induces
throughout optimization. TracIn measures how often and how strongly $z_j$
pushes parameters in directions that also improve performance on an evaluation
point $z$.
Concretely, when $z_j$ appears in a mini-batch $B_t$, TracIn accumulates the
inner product between the gradients at $z$ and $z_j$, weighted by the learning
rate:
\[ 
v_j(z) =
\frac{1}{|B_t|}
\sum_{t : z_j \in B_t}
\eta_t\,
\nabla_\theta \ell(z,\theta_t)^\top
\nabla_\theta \ell(z_j,\theta_t).
\]
In the unified form of Eq.~\eqref{eq:traj_unified}, TracIn corresponds to
setting $\mathcal{B}^{(1)}_t = I$ and letting $\alpha_{t,j}$ be the mini-batch
exposure indicator.

\smallskip

\noindent \textbf{Approach 8: Segmented influence propagation (SOURCE).}
A second class retains closer connection to influence functions, but adapts
them to the discrete and evolving geometry of training. The central observation
by Bae et al.~\cite{bae2024trainingdataattributionapproximate} is that over moderate
timescales, learning dynamics can exhibit approximately stationary curvature
and gradient statistics. SOURCE exploits this by partitioning the trajectory
into segments and approximating gradients and Hessians within each segment by
empirical averages.

For segment $l$, let $\overline{g}_l$ denote the average gradient induced by
$z_j$, $\overline{H}_l$ the average mini-batch Hessian, $K_l$ the segment
length, and $\overline{\eta}_l$ the mean learning rate. Under a linearized
influence approximation to SGD, the contribution accumulated within segment
$l$ is estimated as
\[ 
\overline{r}_{l}
\approx
\frac{1}{n}
\left(I - \exp(-\overline{\eta}_{l} K_{l}\overline{H}_{l})\right)
\overline{H}_{l}^{-1}\overline{g}_{l}.
\]
Crucially, SOURCE propagates earlier contributions forward through later
segments via propagation matrices
$\overline{S}_{l'} \approx \exp(-\overline{\eta}_{l'}K_{l'}\overline{H}_{l'})$,
yielding
\[ 
v_j(z)
=
\nabla_\theta^\top \ell(z,\btheta)
\sum_{l=1}^{L}
\left(
\prod_{l'=L}^{l+1}
\overline{S}_{l'}
\right)
\overline{r}_{l}.
\]

In the unified view of Eq.~\eqref{eq:traj_unified}, SOURCE corresponds to a
piecewise-constant operator $\mathcal{B}^{(1)}_t$ that incorporates both
segment-level inverse curvature and forward propagation along the remaining
training trajectory. This provides a richer temporal model than TracIn, but the
resulting valuation depends on the chosen segmentation and on data-dependent
variation in segment statistics. 

\smallskip
\noindent \textbf{Approach 9: In-run Data Shapley.}
A third trajectory-based line starts from Shapley valuation rather than
influence. Unlike classical Data Shapley, which requires infeasible retraining
on counterfactual subsets, In-run Data Shapley~\cite{wang2024data} assigns
Shapley-style marginal credit directly along the \emph{realized} training
trajectory via stepwise utility increments.
Concretely, let $U_t(\theta_t)$ denote a local utility at step $t$ (e.g., the
validation improvement from $\theta_t$ to $\theta_{t+1}$). The per-step
contribution of $z_j$ is defined as the marginal effect of including $z_j$ in
the update at time $t$, which under a second-order Taylor expansion yields
{
\begin{multline}\label{eq:inrunshap}
\phi_j^{(t)}(z)
\;\triangleq\;
-\eta_t \nabla_\theta \ell(z, \theta_t) \nabla_\theta \ell(z_j,\theta_t) \\
+\frac{\eta_t^2}{2}\nabla_\theta \ell(z_j, \theta_t)^\top H_t^{z}\left(\sum_{z_i \in B_t} \nabla_\theta(z_i, \theta_t)\right),
\end{multline}
}
where $\eta_t$ is the learning rate at step $t$, and $H_t^{z} = \nabla^2_{\theta}(z, \theta_t)$ is the Hessian at step $t$ for validation point $z$. Notice that the first-order-only approximation is equivalent to TracIn formulation where mini-batch size $|B_t|$ being set to $1$.

The overall valuation then aggregates these local credits along
the trajectory,
\[ 
v_j(z)
\approx
\sum_{t=0}^{T-1} \phi_j^{(t)}(z).
\]

In the unified form
Eq.~\eqref{eq:traj_unified}, In-run Data Shapley corresponds to Shapley-weighted exposure coefficients $\alpha_{t,j}$, setting $\mathcal{B}^{(1)}_t = I$ and optional second-order correlation $\mathcal{B}^{(2)}_t[z, D] = H_t^{z}\left(\sum_{z_i \in B_t} \nabla_\theta(z_i, \theta_t)\right)$.

\smallskip

\noindent
\textbf{Summary.}
Eq.~\eqref{eq:traj_unified} highlights that trajectory-aware valuation methods
differ primarily in how they define per-step exposure weights $\alpha_{t,j}$
and propagation operators $\mathcal{B}_t$: TracIn uses raw gradient alignment,
SOURCE introduces segmentwise inverse-curvature and forward propagation, and
In-run Data Shapley incorporates Shapley-style marginal accounting along the
actual training run. While these methods provide scalable attribution beyond
convergence-based influence, their contributions remain trajectory- and
data-dependent accumulations, leaving the worst-case effect of a single
example difficult to bound for differentially private valuation.

\paragraph{Privacy driver (compositional exposure along the trajectory).}
Trajectory-based valuations are characterized by repeated access to the data
across many optimization steps. In Eq.~\eqref{eq:traj_unified}, sensitivity
accumulates through the composed exposure weights and propagation operators,
leading to worst-case dependence scaling with
$\sum_{t=0}^{T-1}\left(\eta_t\left\|\mathcal{B}^{(1)}_t\right\| + \frac{\eta^2_t}{2}\left\|\mathcal{B}^{(2)}_t\right\|\right)$.
This \emph{compositional exposure} is the primary privacy obstacle for
trajectory attribution (Table~\ref{tab:sok_unified_privacy_block}), even when
per-step contributions appear smooth empirically.

%% file: sections/3.4_surrogate.tex
\subsection{Data Modeling and Linearized Attribution}
\label{sec:linear}

A complementary route to scalable valuation replaces the nonlinear training
dynamics of $f(\cdot,\btheta)$ with an intermediate \emph{linear surrogate}
attribution space. A useful unifying template is
\begin{equation}\label{eq:surrogate_unified}
v_j(z)
\;=\;
\mathcal{T}\!\left(\big\langle a(z),\, s(z_j)\big\rangle\right),
\end{equation}
where $a(z)$ is an evaluation-dependent direction (feature query or gradient),
$s(z_j)$ is a training-example embedding in the surrogate space, and
$\mathcal{T}(\cdot)$ maps surrogate similarity into the valuation quantity of
interest. Under this view, contribution is estimated via bilinear similarity
in a tractable linear space rather than via retraining-based marginal effects.

\smallskip

\noindent \textbf{Approach 10: TRAK as a structured Taylor surrogate.}
Park et al.~\cite{park2023trak} provide a canonical instantiation by locally
linearizing the network around the final parameters $\btheta$. For model output
$f(z,\theta)$, TRAK constructs the first-order surrogate
\[ 
\tilde{f}(z,\theta)
\;\approx\;
f(z,\btheta)
+
\nabla_\theta f(z,\btheta)^\top(\theta-\btheta),
\]
so that prediction changes near $\btheta$ are governed by inner products in
parameter-gradient space. This yields influence-style scores that match
Eq.~\eqref{eq:surrogate_unified} with
\[
a(z)=\nabla_\theta f(z,\btheta),
\qquad
s(z_j)=\nabla_\theta \ell(z_j,\btheta),
\]
(optionally preconditioned by a diagonal or Fisher surrogate in practice), and
$\mathcal{T}$ implementing the desired readout (e.g., logit or loss change).
To reduce dependence on a single linearization point, TRAK aggregates scores
across checkpoints or perturbations of $\btheta$, introducing an ensemble
smoothing effect.

\smallskip

\noindent \textbf{Related surrogate variants.}
Other approaches fit naturally into the same template by altering the surrogate
embeddings. \emph{Data models}~\cite{ilyasDatamodelsPredictingPredictions2022}
learn $s(z_j)$ directly through interaction features $\phi(z_j,z)$ and a
predictor $h_\psi$, with $a(z)$ encoded implicitly by the learned map and
$\mathcal{T}$ applied to the surrogate output. \emph{Representation-aware
surrogates}~\cite{ilyas2025magic} instead define $s(z_j)$ in feature space via
intermediate representations $\phi_\ell(z_j)$ and take $a(z)$ as the
corresponding evaluation feature direction. In all cases, valuation reduces to
inner products in a low-dimensional surrogate space, yielding scalability but
requiring additional mechanisms for DP sensitivity control.

\smallskip

\noindent
\textbf{Summary.}
Surrogate-based methods replace nonlinear valuation with bilinear similarity in
an intermediate linear space (Eq.~\eqref{eq:surrogate_unified}). TRAK provides a
structured Taylor instantiation, while data models and representation-based
surrogates offer learned and feature-space variants. These approaches often
smooth valuations empirically, but DP guarantees depend on explicitly bounding
surrogate feature norms rather than on surrogate approximation alone.
\paragraph{Privacy driver (feature-space smoothing without certification).}
Surrogate approaches often reduce variance by projecting attribution into a
lower-dimensional linear space, but DP requires bounding the worst-case
magnitude of surrogate embeddings. Under Eq.~\eqref{eq:surrogate_unified},
sensitivity is governed by quantities such as
$\sup_{z}\|a(z)\|\cdot \sup_{z_j}\|s(z_j)\|$,
which are typically uncontrolled. Thus, surrogate smoothness reflects
generalization bias rather than a dataset-independent sensitivity guarantee,
consistent with Table~\ref{tab:sok_unified_privacy_block}.

%% file: sections/4_takeaways.tex
\section{Lessons Learned}
\label{sec:lessons}

This section synthesizes the analysis of prior sections into a set of
cross-cutting lessons about the feasibility of privacy-preserving data
valuation. Rather than reiterating method-specific details, we focus on the
recurring structural obstacles that arise across valuation families and explain
why they persist despite algorithmic differences. 

\input{sections/4.1_influence_new}

\input{sections/4.2_shap_new}

\input{sections/4.3_traj_new}

\input{sections/4.4_trak}

\vspace{-0.2cm}
\subsection*{Cross-Cutting Challenge} 

\noindent{\bf The Multi-Query Privacy Bottleneck.}
A recurring theme across the three families of valuation methods is the structural mismatch between their single-sample formulation and their multi-sample application. From a computational utility standpoint, there is nothing inherently wrong with defining mechanisms for a \emph{single} record at a time; the process scales trivially by computing each sample independently. However, from a privacy perspective, this scalability is a mirage. Because the total privacy budget $\epsilon$ compounds through composition with every query, the cumulative cost of scoring an entire dataset for curation or auditing quickly becomes prohibitive. Furthermore, as highlighted by previous challenges, most current frameworks fail to meet even baseline \emph{per-record} privacy requirements. The disproportionately high sensitivity of most deep learning utilities demands that the noise required to mask a single individual's data often overwhelms the valuation signal entirely, long before the costs of full-dataset composition are even factored in.

%% file: sections/4.1_influence_new.tex
\subsection{Privacy Challenges in~\autoref{sec:if}}
\label{sec:privacy_challenges}

Despite their algorithmic differences, exact influence functions, iHVP-based approximations (e.g., LiSSA and CG), and Fisher-based surrogates all instantiate the same underlying first-order sensitivity model. The effect of a training point $z_j$ on an evaluation loss is expressed in factorized form as in Eq.~\eqref{eq:if_unified}.
This admits a decomposition into three potentially unbounded components: the evaluation-loss sensitivity $\nabla_\theta \ell(z,\btheta)$, the training-point contribution $\nabla_\theta \ell(z_j,\btheta)$, and the inverse-curvature operator $\mathcal{A}_\lambda$.

\smallskip
\noindent{\bf C1. The heavy-tail phenomenon: geometry and distribution.}
\begin{figure}[t]
    \centering
    \includegraphics[width=0.9\linewidth]{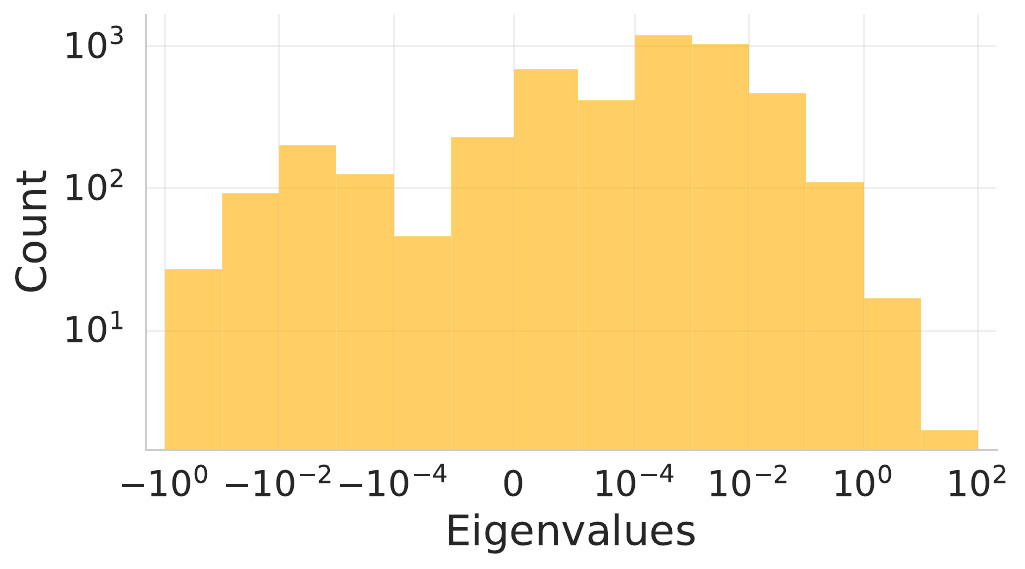}
    \caption{\textbf{Spectral distribution of the empirical Hessian.} The distribution of eigenvalues $\mu$ for the empirical Hessian $H$ of a CNN trained on MNIST. The high concentration of eigenvalues with magnitude near zero is consistent with the property of flatness in the loss landscape at convergence. This implies that the operator norm $\|H^{-1}\|$ is difficult to bound and potentially ill-defined. \vspace{-4mm}}
    \label{fig:h_spectrum}
\end{figure}
The central challenge for privacy lies in the interaction between the loss landscape geometry and the resulting distribution of influence scores. In modern neural networks, the empirical Hessian $H = \tfrac{1}{n}\sum_{i=1}^n \nabla_\theta^2 \ell(z_j,\btheta)$ is typically ill-conditioned. Stochastic optimization frequently converges to regions of the loss landscape that are ``flat'' in many directions. In such regimes, $H$ contains eigenvalues near zero, causing the inverse operator $\mathcal{A}_\lambda = H^{-1}$ to exhibit correspondingly large eigenvalues. These extreme eigenvalues act as a magnifier for specific gradient directions.
We empirically demonstrate this using a small CNN trained on MNIST. As shown in \autoref{fig:h_spectrum}, the majority of the eigenvalues of the Hessian $H$ lie between $-10^{-2}$ and $10^{-2}$. When the inverse-curvature operator is applied, training gradients that align with eigenvectors corresponding to these near-zero eigenvalues are amplified by orders of magnitude.

\begin{figure}[t]
    \centering
    \includegraphics[width=0.9\linewidth]{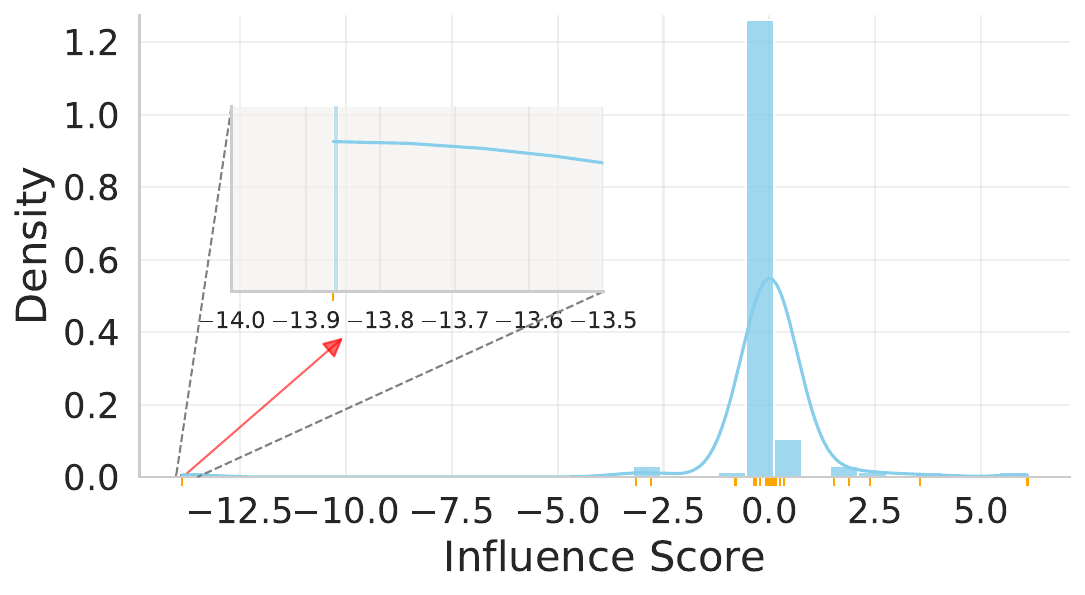}
    \caption{\textbf{Distribution of influence scores.} Influence scores for 100 sampled training data points on a fixed validation point. Most influence scores are centered around zero, but extreme outliers exist due to curvature amplification. The presence of these outliers impacts sensitivity estimation.\vspace{-4mm}}
    \label{fig:if_score}
\end{figure}

This geometric instability results in a \textit{highly skewed, heavy-tailed distribution} of influence scores. As visualized in \autoref{fig:if_score}, the vast majority of scores are clustered tightly around zero. However, a small subset of ``impactful'' outliers--whose gradients align with the flat directions of the Hessian--possess scores with magnitudes significantly larger (e.g., $10^1$ versus $10^{-5}$). From a privacy perspective, this skewness is problematic: the sensitivity is dictated by the extreme outliers, while the utility is derived from the granular ranking of the majority.

\smallskip
\noindent{\bf C2. Unbounded sensitivity and the limits of damping.}
Because the influence score relies on the inverse Hessian, the global sensitivity is theoretically unbounded without modification. A common strategy to stabilize computation is damping, where the inverse operator is modified to $\mathcal{A}_\lambda = (H + \lambda I)^{-1}$. This can also be seen as a term rising naturally from the loss function with $2$-norm regularization where $\ell'(z, \theta) = \ell(z, \theta) + \frac{\lambda}{n} \|\theta\|^2$.

\noindent \underline{\em The Tuning Dilemma.}
Previous studies have demonstrated that the choice of $\lambda$ significantly impacts the predictive quality of influence scores~\cite{basu2021influence, wang2025taminghyperparametersensitivitydata}. There typically exists different values of an ``optimal'' $\lambda$ that maximizes valuation performance for different tasks and models (e.g., for mislabeling detection), but this value is dataset-dependent and requires tuning. Crucially, the $\lambda$ that maximizes utility is rarely the same as the $\lambda$ required to strictly bound the sensitivity for DP.

\noindent \underline{\em Failure in Indefinite Regimes (Hessian).}
When using the exact Hessian or non-PSD approximations (such as those derived from Arnoldi iteration or random projections), $H$ is indefinite. As shown in \autoref{fig:h_spectrum}, the Hessian spectrum extends into negative values. In this regime, damping only guarantees a bound if $\lambda > |\lambda_{\min}(H)|$. If the performance-tuned $\lambda$ is smaller than the magnitude of the negative eigenvalues, the matrix $H + \lambda I$ remains indefinite or singular. Consequently, the inverse operator $(H + \lambda I)^{-1}$ remains unbounded, providing no sensitivity guarantee whatsoever.

\noindent \underline{\em Loose Bounds in PSD Regimes (Fisher).}
Even when using positive semi-definite (PSD) surrogates like the Fisher Information Matrix (FIM) $\hat{H} \succeq 0$ and damping guarantees positive eigenvalues $\ge \lambda$--the resulting bound is practically insufficient. The operator norm is bounded by $\|(H+\lambda I)^{-1}\| \le 1/\lambda$. Since $\lambda$ is typically small (e.g., $10^{-1}$ to $10^{-3}$)~\cite{basu2021influence, wang2025taminghyperparametersensitivitydata} to preserve the curvature signal, the resulting sensitivity bound $1/\lambda$ is extremely loose. Relying on this worst-case bound for noise addition would require injecting noise magnitudes larger than the signal of the influence scores themselves.

Thus, while damping suppresses some pathological curvature directions, it does not yield a tight, dataset-independent sensitivity bound usable for privacy mechanisms.

\begin{figure}[t]
    \centering
    \includegraphics[width=0.9\linewidth]{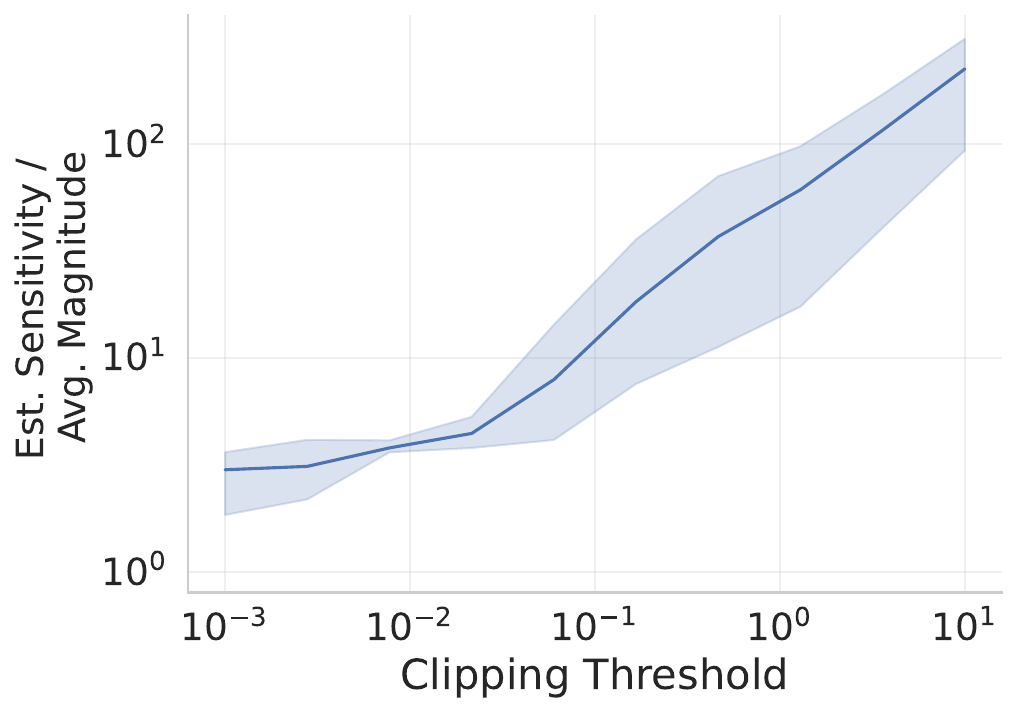}
    \caption{\textbf{Ratio of estimated sensitivity to average score magnitude.} Even with tight clipping, the noise required (proportional to sensitivity) overpowers the signal of the average data point. The ratio remains $>1$, indicating that the noise floor exceeds the signal for the majority of the distribution.\vspace{-4mm}}
    \label{fig:rel_sens}
\end{figure}

\smallskip
\noindent{\bf C3. The privacy vs. utility paradox in point-wise release.} To enforce finite sensitivity, one must explicitly bound the magnitude of the components via clipping. By clipping the gradients (of both training and evaluation points) to a threshold $\gamma$, we can derive a global sensitivity bound. For a vector $v$, let $\mathrm{clip}_\gamma(v) = \min\{1, \gamma/\|v\|\} \cdot v$. We can then compute clipped estimates in different ways. For example, by transposing the computation in Eq.~\eqref{eq:if_unified}, we can enforce sensitivity by clipping the intermediate matrix-vector re:
\begin{equation}
    \hat{v_j}(z) = -\mathrm{clip}_\gamma(\nabla_\theta \ell(z_j,\btheta))^\top \mathrm{clip}_\gamma\left(\mathcal{A}_\lambda \nabla_\theta \ell(z,\btheta)\right)
\end{equation}

\noindent \underline{\em The Distortion of Tail-Heavy Distributions.} 
While clipping provides a theoretical bound, it creates a fundamental conflict with the point-wise nature of influence score release. Unlike aggregate mechanisms (e.g., DP-SGD) where noise is averaged over a batch, influence functions release a distinct score for every training point $z_j$.

\begin{squishenumerate}
    \item {\em Loss of Outlier Resolution (Clipping too tight):} If $\gamma$ is set low to match the majority of the data (the ``middle bunch''), the impactful outliers are clipped to $\gamma$. They become indistinguishable from any other point that hits the threshold, effectively erasing the information that they were orders of magnitude more influential.
    
    \item {\em Noise Overwhelms the Majority (Clipping too loose):} If $\gamma$ is set higher to preserve the relative magnitude of the outliers, the global sensitivity increases proportionally. Since DP noise scales with the global sensitivity, the added noise becomes orders of magnitude larger than the true scores of the majority.
\end{squishenumerate}
    
    As demonstrated in \autoref{fig:rel_sens}, even with aggressive clipping, the ratio of the estimated sensitivity to the average score magnitude remains greater than 1. Consequently, any noise mechanism sufficient to satisfy DP will have a magnitude comparable to or larger than the scores themselves, rendering the released values useless for the majority of the dataset. This highlights that simply enforcing finite sensitivity via clipping is insufficient for the private release of fine-grained, tail-heavy influence scores.

\smallskip
\noindent{\bf Design takeaway.}
The failure of influence-based valuation under DP is structural rather than algorithmic. %
The aforementioned observations imply that privacy-friendly data valuation cannot be achieved by retrofitting DP onto influence scores, but instead requires rethinking the valuation objective itself—favoring locality, bounded interaction scopes, or structurally constrained statistics that limit worst-case influence by construction rather than by post hoc regularization.

%% file: sections/4.2_shap_new.tex
\subsection{Privacy Challenges in~\autoref{sec:marginal}}
Data Shapley, Beta Shapley, and Data Banzhaf differ primarily in how these
subsets are weighted and sampled, but they share the same conceptual backbone:
a data point is valuable insofar as it improves model performance \emph{on
average across hypothetical retraining scenarios}. As a result, we take Data Shapley as a demonstration of privacy challenges and potential solutions in weighted marginal contribution schemes.

\smallskip
\noindent{\bf C4. Disproportionally large empirical sensitivity for common utility functions.}
The computation of Shapley values hinges on the marginal contribution of a data point, defined as the difference in utility $\Delta_{S, z_j}$ when $z_j$ is added to a subset $S$. For neural networks, the utility function $U$ is typically accuracy or loss. However, deep learning models lack tight theoretical bounds on how the performance changes when a single training point is perturbed. This shows up in particular when the marginal value samples small subsets $S$, mimicking the volatile early stages of training where utility improvements are rapid and unstable.
To quantify this, we trained a small fully connected network on the \texttt{scikit-learn digits} dataset. We estimated Shapley values using several standard approximation methods\footnote{Using \texttt{pyDVL} package for the implementation, available at
\url{https://github.com/aai-institute/pyDVL}.} introduced in \autoref{sec:marginal}--naive MC, stratified, permutation, and truncated MC Shapley--on both the original dataset and a modified version with a single perturbed training point, using accuracy as utility function. We record the maximal change in Shapley scores as a lower-bound estimation of global sensitivity. \autoref{tab:shapley_sensitivity} directly demonstrates that hat even under optimistic empirical conditions, the noise required to mask the presence of a data point would likely overwhelm the signal.

\begin{table}[t]

    \centering
    \begin{tabular}{l|c}
        \toprule
        \textbf{Method} & \textbf{Est. Sensitivity / Avg. Magnitude} \\
        \midrule
        Naive MC $(N=5000)$ & > 100 \\
        Stratified (Uniform $k$) & 4.947 \\
        Stratified (Fixed $k$) & 1.009 \\
        Permutation & 4.597 \\
        Truncated MC & 4.699 \\
        \bottomrule
    \end{tabular}
    \caption{\textbf{Ratio of estimated sensitivity to average score magnitude.} This records the maximal changes relative to the average score magnitude in Data Shapley scores across \texttt{scikit-learn digits} dataset (split 80\%-20\% for training and validation) upon flipping a single label in the dataset.
    We observe that the empirical sensitivity often matches or exceeds the magnitude of the scores themselves.}
    \label{tab:shapley_sensitivity}
\end{table}

\smallskip
\noindent{\bf C5. Aggregation vs. sensitivity trade-off remains hard to balance.}
To formally control sensitivity, we can apply clipping, $\operatorname{clip}_\gamma(\Delta_{S, z_j})$, ensuring each marginal contribution is bounded by $2\gamma$. A clipped Data Shapley value can be computed as follows:
\[ 
v'_j = \sum_{k=1}^{n} \frac{1}{n} \binom{n-1}{k-1}^{-1} \sum_{\substack{S \subseteq D \setminus \{z_j\},\\ |S|=k-1}} \operatorname{clip}_\gamma(U(S \cup \{z_j\}) - U(S)).
\]
In standard private aggregation (e.g., mean estimation), aggregation reduces sensitivity because a single user affects only one term in the sum. The unique challenge in Data Shapley is that a single training point $z_j$ participates in multiple subsets.

In exact Shapley computation, $z_j$ appears in exactly half of all $2^{ |D| -1}$ possible subsets. Even if each marginal term is clipped to $\gamma$, the global sensitivity accumulates linearly with the number of terms involving $z_j$.
To satisfy DP, we must limit the global sensitivity. This creates a direct trade-off involving the approximation method:
\begin{squishenumerate}
    \item \em{High Sampling Rate (High Utility, High Sensitivity):} To get a stable, low-variance estimate of the Shapley values, we generally use permutation based estimations~\cite{Mitchell_Cooper_Frank_Holmes_2022}. While some early stopping may apply when iterate over the permuted dataset, this still increases the probability that the perturbed point $z_j$ is included in the background sets, causing it to affect more marginal terms. This inflates the global sensitivity, necessitating higher noise injection.
    \item \em{Low Sampling Rate (Low Utility, Low Sensitivity):} We can algorithmically limit the number of times any specific data point $z_i \neq z_j$ is allowed to participate in a background set $S$ globally. While this successfully bounds the sensitivity (limiting the ``fan-out'' of $z_j$'s influence), it restricts the total number of subset samples available to estimate the Shapley values for the rest of the dataset.
\end{squishenumerate}

Watson et al.~\cite{watson2022differentiallyprivateshapleyvalues} propose a DP algorithm for cases where the marginal term is naturally bounded (avoiding the bias of clipping). However, to keep the sensitivity manageable, their method results in an estimator with performance comparable only to Naive MC, which is widely considered a poor estimator with slow convergence compared to methods like permutation sampling~\cite{Mitchell_Cooper_Frank_Holmes_2022, goldwasser2024stabilizingestimatesshapleyvalues}.

In conclusion, applying DP to Data Shapley encounters a critical bottleneck at the model training step. As detailed in \autoref{sec:marginal}, the utility instability of deep learning models renders the global sensitivity effectively unbounded. This forces reliance on loose, worst-case noise levels that overwhelm the subtle signals required for fine-grained data valuation.

\smallskip
\noindent{\bf C6. Sensitivity by design is hard to achieve.}
Given the trade-off imposed by the aggregation structure, Wang et al.~\cite{wang2023privacy} provide an alternative path via the Threshold $k$-NN (T$k$-NN) Shapley framework (details in~\autoref{sec:locality}). 
T$k$-NN achieves bounded sensitivity \textit{by design} through a ``noise floor'' mechanism. If a query point is too far from the training data (exceeding a distance threshold), the utility function reverts to random guessing. 
This structural bounding offers a robust alternative to the high-variance reality of standard deep learning utilities described in \textbf{C4}, effectively circumventing the disproportionally large marginal contributions that make raw utility metrics difficult to privatize.
This suggests that for deep networks, achieving DP Shapley values may require redefining the utility $U$ to inherently limit the influence of outliers (sensitivity by design), rather than relying on clipping or subsampling (sensitivity by post-processing) which forces a compromise on estimation quality.

\smallskip
\noindent{\bf Design Takeaway.} The 
The most effective way to enable privatization is to establish structural stability by design instead of applying post-hoc correction: rather than attempting to privatize a fundamentally volatile utility function (like raw accuracy) through aggressive clipping, focus on "sensitivity-by-design" architectures. This involves using utility functions that are naturally Lipschitz-continuous or locality-constrained (e.g., T$k$-NN or kernel-based methods), which provide a predictable noise floor. For utility that is inherently sensitive, it remains difficult to balance the subset coverage (which ties to quality of the score) with the accumulation of sensitivity.

%% file: sections/4.3_traj_new.tex
\subsection{Privacy Challenges in~\autoref{sec:traj}}
Compare to influence- and marginal-based methods, trajectory-based methods evaluate data by auditing the actual optimization path taken by the model. This shift in perspective offers a unique privacy advantage: if the training process itself is conducted under a DP guarantee, the valuation can be treated as a "free" post-processing step. Crucially, however, the privacy-friendliness of these methods is strictly contingent on the use of DP-optimizers like DP-SGD; in the absence of a private trajectory, these methods offer no inherent protection.

\smallskip
\noindent{\bf C7. Limited DP-compatibility of trajectory-based methods.}
Under standard privacy accounting (e.g., the moments accountant), the DP-SGD algorithm guarantees that the entire trajectory of model checkpoints $\Theta = [\theta_1, \dots, \theta_T=\boldsymbol{\theta}]$ constitutes a differentially private release~\cite{Abadi_Chu_Goodfellow_McMahan_Mironov_Talwar_Zhang_2016}. 

\noindent \underline{\em Direct DP Adoption of First-Order Methods.}
TracIn's formulation (and consequently, in-run Shapley's first order form) is privacy-friendly in the sense that no other training data is explicitly involved at the time of computing valuation for a given sample. Consequently, the first-order TracIn calculation $v_j(z)=\sum_{t} \nabla \ell(z, \theta_t)^\top \nabla\ell(z_j, \theta_t)$ functions strictly as a post-processing operation $f(\Theta, z_j, z)$, ensuring that influence scores $v_j(z)$ satisfy $(\epsilon, \delta)$-DP with respect to the remaining training set $D \setminus \{z_j\}$. 

\noindent \underline{\em Modifying Second-Order Methods.}
The second-order in-run Shapley introduces an extra layer of interaction term involving the summation of gradients from the entire mini-batch $\sum_{z_k \in B_t} \nabla \ell(z_k)$. While this term mathematically depends on the sensitive mini-batch data $B_t$, privacy can be preserved by substituting the raw gradient sum with the \textit{observed} update vector from the released trajectory $\Theta$ (which essentially represents the noisy, clipped sum). By computing the second order interaction in Eq.~\eqref{eq:inrunshap} as $\nabla_\theta \ell(z_j, \theta_t)^\top H_t^z (\theta_{t+1} - \theta_t)$, the calculation remains a function of public/released variables, maintaining the post-processing guarantee. Provided that the validation data $z$ is public (or synthetic) and $z_j$ is known to the user, no re-access to the raw private database is required.

\begin{figure}[t!]
    \centering
    \includegraphics[width=\linewidth]{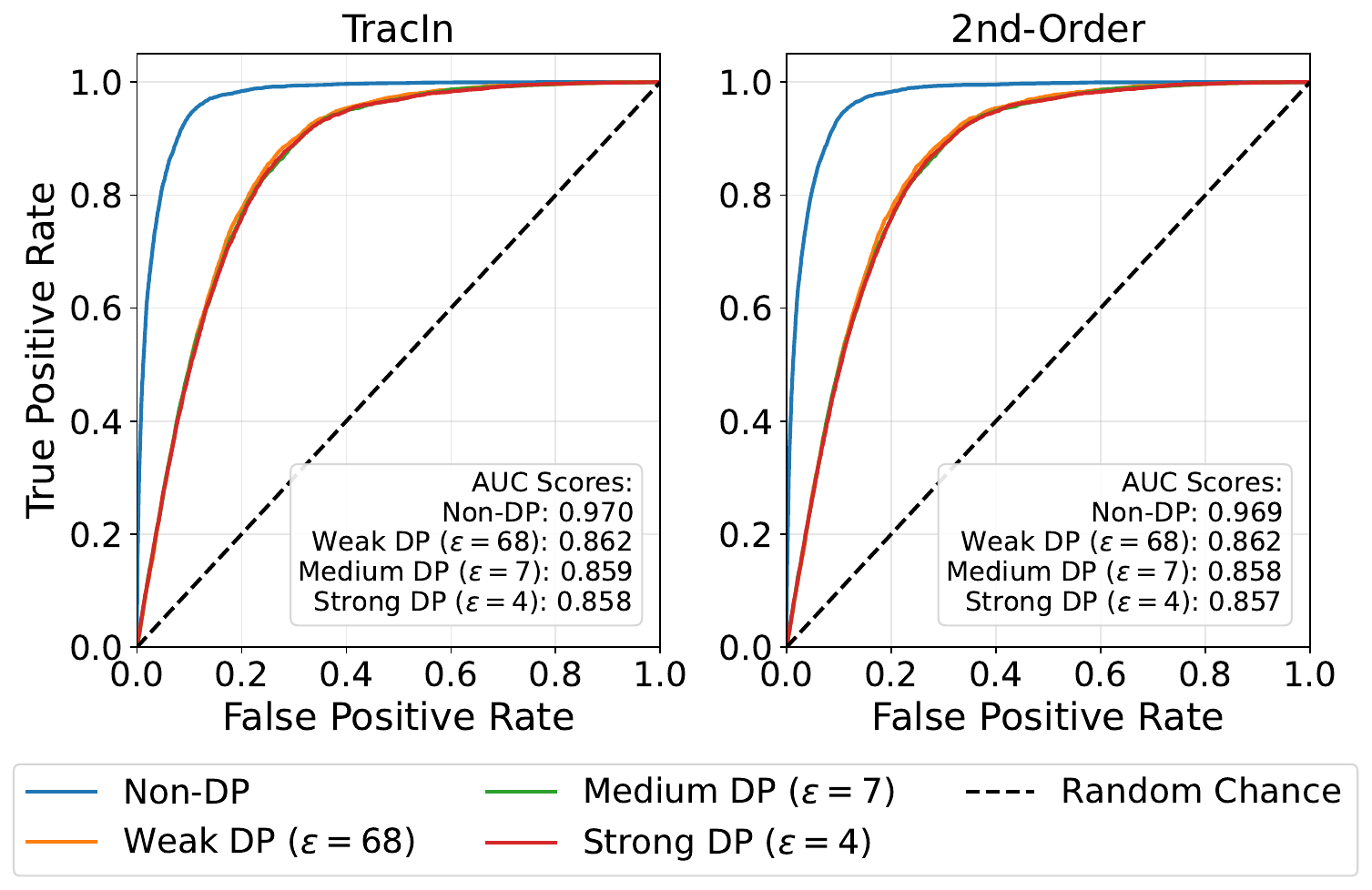}
    \vspace{-.3cm}
    \caption{ROC curves evaluating mislabel detection performance across different privacy budgets. Results are very similar for both the TracIn and the 2nd-order in-run Shapley methods. With regular training, the valuation methods are very successful in detecting mislabeled data, while DP-SGD degrades the performance by a slight amount. The strength of noise level in DP noise negatively affects mislabel detection performance very slightly.}
    \label{fig:mislabel}
    \vspace{-.3cm}
\end{figure}

\noindent \underline{\em Experimental Investigation of Privacy-Utility Tradeoff:} We train a small convolutional neural network with 3 convolutional layers and around 600k parameters on CIFAR10 with and without DP-SGD, and compute the following valuation task:

\begin{squishenumerate}
    \item Top-$k$ overlap: we randomly sampled testing data $z$'s. For each sample $z$ in the set, we compute the TracIn and modified second order in-run Shapley scores for all training data and choose the top-$k$ training samples $S$ that have the highest scores. For different noise multiplier $\sigma$ and resulting privacy budget $\varepsilon$, we compute the percentage of the overlap between the choices of model trained with and without differential privacy.

    \emph{Note}: since the data valuation scores is also sensitive to training trajectories and hyperparameters~\citet{wang2025taminghyperparametersensitivitydata}, the different hyperparameters for DP-SGD from regular training can contribute the mismatch in top-$k$ selections. As a result, we compare models trained with DP-SGD with model trained with clipping but no noise mechanism such that the hyperparameters are consistent throughout the model training runs.

    As shown in \autoref{tab:topk_overlap}, for the first order method (TracIn), model trained with DP suffers a hit in performance, where even model trained with very weak DP ($\varepsilon=68$) only managed to have around 50\% overlap with model trained without noise. For the modified second-order in-run Shapley method, the extra noise used results in a utility degradation on a similar level, with models achieving around 40\% overlap regardless of the privacy budget.

    \item Mislabel detection:~\citet{pruthi2020estimatingtrainingdatainfluence} used mislabel detection via self-influence i.e. setting $z_j = z$ as evaluations for data valuation effectiveness. Mislabeled data would usually have higher loss and have a stronger effect on the model behavior, thus having a high score~\citep{pruthi2020estimatingtrainingdatainfluence}. The model is trained on corrected data, and then 10\% of the training are manually perturbed for detection.

    By setting different score threshold for mislabel detection, we can compute the AUC (area under curve) between DP and non-DP methods.

    As \autoref{fig:mislabel} shows, although privacy-utility tradeoff exists, here models trained with DP-SGD still performs reasonably well in mislabel detection for small models.
\end{squishenumerate}

\begin{table}[t]
\centering

\begin{tabular}{l cc}
    \toprule
    & \multicolumn{2}{c}{\textbf{Top-$100$ Overlap Rate}} \\
    \cmidrule(lr){2-3}
    & \textbf{TracIn} & \textbf{2nd-order} \\
    \midrule
    Weak DP ($\sigma=0.5, \varepsilon=68$)  & 0.522 & 0.393 \\
    Medium DP ($\sigma=1.5, \varepsilon=7$) & 0.461 & 0.411 \\
    Strong DP ($\sigma=2.5, \varepsilon=4$) & 0.471 & 0.406 \\
    \bottomrule
\end{tabular}
\caption{Top-$100$ data valuation overlap between non-DP (trained with clipping) and DP models across varying privacy budget ($\varepsilon$) computed by TracIn and 2nd-order in-run Shapley. Overlap is measured as percentage of the size of the exact intersection of the most influential training examples.}
\label{tab:topk_overlap}
\vspace{-.2cm}
\end{table}

\noindent \underline{\em Trajectory's Limits.}
It is important to note, however, that this reliance on trajectory privacy precludes the use of ``hidden state'' privacy amplification techniques~\cite{Feldman_Mironov_Talwar_Thakurta_2018, Altschuler_Talwar_2023, Ye_Shokri_2022, Chien_Li_2025}, as the intermediate checkpoints and update vectors must be explicitly utilized for the valuation.

Despite providing privacy guarantee for a single point release $v_j(z)$, these methods are still unsuitable for the \textit{central release} model (dataset-level outputs). The ``post-processing'' privacy argument holds only when the valuation function $v(\Theta, z)$ is applied to a \textit{public} or user-provided query point $z$, ensuring that the only dependence on the private database $D$ is through the sanitized trajectory $\Theta$. However, in a central release scenario, the goal is to publish the vector of influence scores $V = \{v(\Theta, z_i) \mid z_i \in D\}$ for the training data itself. In this context, the valuation function requires direct access to the raw private training points $z_i$ to compute their individual gradients $\nabla \ell(z_i)$.

\smallskip
\noindent{\bf C8. Other schemes still suffer from compounding sensitivity.}
While first-order methods or those utilizing public second-order information align with trajectory-based privacy, alternative schemes relying on the Hessian of the \textit{training} data still face severe privacy hurdles. Methods such as the SOURCE framework~\cite{bae2024trainingdataattributionapproximate} explicitly model the optimization path using the curvature of the private loss landscape $H$. While SOURCE shows great performance advantage over TracIn, this dependence creates a problem of compounding sensitivity. Even if the model updates themselves are privatized via DP-SGD, the curvature information used by these methods cannot be directly computed from the trajectory $\Theta$. Rather, calculating attributions using $\overline{H}_{l}$ constitutes a fresh, non-private query to the raw training batches, violating the post-processing assumption.

\smallskip
\noindent{\bf Design takeaway.} Trajectory-based attribution methods exhibit a distinct privacy constraint. First-order methods (like TracIn) are uniquely amenable to DP as they function as post-processing operations on the released DP-SGD trajectory. However, this compatibility imposes two strict limits: 
\begin{squishenumerate}
    \item It precludes the use of advanced ``hidden state'' privacy amplification accounting (which improves model utility by hiding intermediate checkpoints), as the explicit trajectory is required for valuation. %
    \item It prohibits the use of private second-order training information (the training Hessian), as this data is not protected by the standard gradient noise mechanism.
\end{squishenumerate}

%% file: sections/4.4_trak.tex
\subsection{Privacy Challenges in~\autoref{sec:linear}}
Surrogate methods serve as a conceptual bridge between the marginal-utility frameworks of~\autoref{sec:marginal} and the trajectory-based audits of~\autoref{sec:traj}. On one hand, they inherit the \textit{counterfactual} goal of marginal methods, seeking to approximate how a model's performance would change if a specific subset $S$ were modified. On the other hand, they utilize the \textit{local curvature information} of influence-based methods to bypass the prohibitive cost of actual model retraining. However, it also inherits privacy challenges from both sides.

\smallskip
\noindent{\bf C9. Hidden global dependence in surrogate geometry}
The linear surrogate template $v_j(z) = \mathcal{T}(\langle a(z), s(z_j) \rangle)$ superficially suggests a clean, pairwise isolation between the query $z$ and the training point $z_j$. However, this modularity is often an illusion. In high-utility instantiations like TRAK or influence functions, the mapping functions $a(\cdot)$ and $s(\cdot)$ are not independent of the remaining dataset. They typically incorporate a \textit{preconditioning matrix} $P$ such as the inverse Hessian $H^{-1}$ or the Empirical Fisher Information Matrix ${G}^{-1}$---to correct for the curvature of the loss landscape:
\[
s(z_j) \;=\; {P}(D) \cdot \nabla \ell(z_j).
\]
This matrix ${P}(D)$ is an aggregate statistic derived from the \textit{entire} private dataset $D$ (or large batches thereof). Consequently, constructing the surrogate embedding $s(z_j)$ for a single user requires a fresh, global query to the sensitive database to compute the curvature. As noted in \textbf{C2/C8}, the sensitivity of this matrix is often unbounded or effectively high-dimensional, preventing compliant privatization via standard noise mechanisms. Thus, even if the gradient $\nabla \ell(z_j)$ is handled carefully, the geometry of the space it lives in leaks information about $D \setminus \{z_j\}$.

Furthermore, the ensemble techniques used to stabilize these surrogates (e.g., averaging scores over multiple trained model on different subsets $S_1, \ldots,S _M$ in TRAK) offer little privacy benefit for the background data $D \setminus \{z_j\}$. As mentioned in \textbf{C5}, a single training data can appear in multiple ensembles. As a result, in the perturbation of a data point can have multiple copies of influence in the scores, limiting the effect of aggregation.

\smallskip
\noindent{\bf Design Takeaway.} The challenges identified in the surrogate framework suggest that privacy-preserving valuation may require moving away from purely data-dependent geometries. Possible design directions include:
\begin{squishenumerate}
    \item {\em Data-independent geometries}: One potential path to mitigate geometric leakage is the use of projections that do not depend on the private training set (e.g. public-data priors). While this may sacrifice some alignment with the true loss landscape, it offers a path to computing embeddings $s(z_j)$ that are structurally decoupled from the rest of the sensitive dataset.
    \item {\em Managing participation in ensembles}: Similar to Shapley-based methods, methods such as TRAK utilizes ensembling over subsets to improve score stability and performance. As such, it remains a critical task to establish better tradeoff between subset coverage and sensitivity accumulation.
\end{squishenumerate}

%% file: sections/5_open.tex
\section{Open Problems}
\label{sec:open}

The analysis of challenges {\bf C1} through {\bf C9} reveals a fundamental tension: the geometric information required for high-utility valuation (curvature, subsets, ensembles) is often the exact source of privacy leakage. Resolving this requires moving beyond standard DP-SGD analysis. We highlight three critical directions for future research.

\smallskip
\noindent{\bf P1. Tighter accounting for trajectory-based valuation.}
As noted in \textbf{C7}, first-order trajectory methods (e.g., TracIn) currently offer the most promising path for DP compliance because they satisfy the post-processing property of the optimization path $\Theta$. However, this creates a conflict with modern privacy accounting. State-of-the-art DP-SGD accountants (e.g., for hidden state amplification) rely on the assumption that intermediate checkpoints are \textit{not} released.
Data valuation inverts this requirement: it demands high-resolution access to the trajectory (dense checkpoints) to integrate the gradient alignment over time.
\begin{squishitemize}
    \item \textit{The challenge:} How can we derive tight privacy bounds for the explicit release of the \textit{gradient alignment sequence} used in valuation, rather than the full high-dimensional trajectory?
    \item \textit{Future direction:} Developing specialized ``Valuation Accountants'' that quantify the privacy cost of releasing scalar products $\langle \nabla \ell(z), \nabla \ell(z_j) \rangle$ over time. Additionally, since second-order information (Hessian) is effectively forbidden (\textbf{C8}), research could focus on boosting the utility of first-order approximations—perhaps via safe public-data preconditioning—to match the performance of prohibited second-order methods like SOURCE.
\end{squishitemize}

\smallskip
\noindent{\bf P2. Is static (final-model) task-agonistic DP valuation possible?}
Challenges \textbf{C1}, \textbf{C4}, and \textbf{C9} present a grim picture for valuation methods that rely solely on the final model state or static retraining outputs.
For Influence Functions, the bottleneck is the reliance on the private inverse Hessian $H^{-1}$ to disentangle feature correlations (\textbf{C1}, \textbf{C2}).
Similarly, for Data Shapley, the barrier is the \textit{unbounded marginal utility} of deep neural networks (\textbf{C4}). Existing DP Shapley algorithms~\cite{watson2022differentiallyprivateshapleyvalues} rely on bounded utility assumptions that do not generally hold neural networks, while structurally bounded alternatives like T$k$-NN~\cite{wang2023privacy} are restricted to specific local tasks and do not generalize to the full deep learning setting.
\begin{squishitemize}
    \item \textit{The challenge:} Can we extract meaningful attribution from a converged, private model without accessing the private curvature $H$ or exposing the highly sensitive marginal utility of the training data?
    \item \textit{Future direction:} Investigating \textit{Public-Data Surrogates}. If the loss landscape geometry or the marginal utility distribution can be adequately approximated using a public dataset (assuming with adequate utility), one could compute the preconditioner $H_{pub}^{-1}$ or estimate Shapley proxies safely. The open question is whether public-data curvature is sufficiently aligned with private-data curvature to yield accurate attribution for outliers, or if the fitted ``flat'' regions of the landscape (\textbf{C1}) make such transfer impossible.
\end{squishitemize}

\smallskip
\noindent{\bf P3. Privacy beyond per-record release: central and two-way leakage.}
We have mostly discussed the per-record release threat model: a single, private training point $z_j$ being evaluated against a \textit{public} validation sample $z$, with the score released to the data owner. Real-world deployment scenarios break this model in two distinct ways, creating unsolved privacy bottlenecks.
\begin{squishitemize}
    \item \textit{The central release problem (Leakage of $D_{train}$):} Some scenarios require publishing the valuation vector $V$ for the whole or a large portion of the training set. This creates a problem of scale that current methods--which already struggle to privatize a single score $v_j(z)$--are ill-equipped to handle. Naive composition of per-record mechanisms would deplete the privacy budget immediately, and no current mechanism exists to efficiently bound this global vector sensitivity.
    \item \textit{The private validation problem (Leakage of $D_{val}$):} Current methods assume the validation set $z_{val}$ is public. In many sensitive domains (e.g., healthcare federation), the validation set belongs to a distinct party (e.g., a hospital) and is itself private. Releasing the influence score $v_j = \langle \nabla \ell(z_j), \nabla \ell(z_{val}) \rangle$ to the training data owner $z_j$ mathematically reveals the gradient of the validation data. This creates a \textit{two-way leakage} problem where both the training data and the audit query must be simultaneously protected.
    \item \textit{Future direction:} Addressing these issues requires moving beyond single-sided, scalar DP. For dataset-level release, research could focus on \textit{high-dimensional output perturbation} or leveraging \textit{algorithmic stability} to bound the sensitivity of the global ranking. For private validation, the problem can utilize \textit{Secure Multi-Party Computation (SMPC)}~\cite{4568388} or homomorphic encryption as well as differential privacy, where the goal is to compute the influence-style inner product of gradients without revealing the raw vectors to either party.
\end{squishitemize}

%% file: sections/6_conclusion.tex
\section{Conclusions}
\label{sec:conclusion}

This SoK reveals that the conflict between data valuation and privacy is not merely an accounting inconvenience, but potentially a structural contradiction. The core signal of valuation—the fine-grained sensitivity of model behavior to a specific record—is precisely the leakage that differential privacy seeks to suppress. Our analysis demonstrates that widely used valuation paradigms fail to resolve this tension: Shapley-based methods suffer from the unbounded marginal utility of deep networks (\autoref{sec:marginal}), influence functions rely on prohibited private curvature (\autoref{sec:if}), and surrogate ensembles exhibit massive global coupling (\autoref{sec:linear}).

Consequently, retrofitting standard DP mechanisms like output perturbation or clipping onto existing algorithms proves ineffective. This prompts a reconsideration and redesign in data valuation schemes. Ultimately, meaningful privacy in data valuation will not come from silencing the influence of data, but from designing valuation mechanisms that can disentangle the \textit{valid signal} of data quality from the \textit{private signal} of individual identity.

%% file: sections/app0.tex
\section{Detailed Interpretation of~\autoref{tab:sok_unified_privacy_block}}
\label{app:table_explanation}

Table~\ref{tab:sok_unified_privacy_block} provides a privacy-oriented
systematization of data valuation methods. Rather than organizing approaches
purely by computational technique (e.g., Hessian approximations versus subset
sampling), the table highlights the \emph{structural sensitivity drivers} that
determine whether valuation scores can be released with meaningful worst-case
privacy guarantees. Each row corresponds to one estimator family surveyed in
Sections~\S\ref{sec:if}--\S\ref{sec:locality}, and each column captures a
property that governs amplification, composition, and DP feasibility.

\paragraph{Symbol semantics.}
Entries are qualitative: \Full\ indicates that the property is strongly present
or explicitly enforced by the estimator design; \Half\ indicates partial or
heuristic presence, often dependent on modeling assumptions, truncation, or
implementation details; \Empty\ indicates absence. Importantly, these ratings
are not intended as absolute judgments of method quality, but as a compact
summary of which \emph{privacy-relevant mechanisms} dominate in each family.

\medskip
\noindent\textbf{Structural columns.}
The first block characterizes how valuation is computed and where sensitivity
enters the estimator.

\begin{squishitemize}

\item \textbf{Retrain.}
This column indicates whether valuation is defined through repeated retraining
on counterfactual datasets. Coalition-based methods such as Data Shapley,
Beta Shapley, and Data Banzhaf score \Full\ because their values are defined as
expected marginal utility differences over subsets, which in principle require
training a model on many coalitions. In contrast, influence-based and
surrogate-based approaches avoid retraining by working with a single trained
model and local approximations, and therefore score \Empty. Retraining matters
for privacy because it introduces repeated data access and makes sensitivity
accounting substantially more complex.

\item \textbf{Exposure.}
Exposure captures the degree of \emph{compositional access} to the private
dataset required by the estimator. In DP, privacy loss accumulates under
composition when a mechanism makes repeated data-dependent queries. Trajectory
methods (TracIn, SOURCE, In-run Data Shapley) score \Full\ exposure because
their valuations explicitly sum or propagate contributions across many training
steps, effectively composing sensitivity over $T$ iterations. Coalition-based
methods also incur \Full\ exposure since estimating marginal contributions
requires evaluating many subsets or permutations. By contrast, classical
influence approximations are closer to a single-shot computation at the final
iterate and thus score \Empty. Locality-based methods achieve intermediate
(\Half) exposure because their computations involve repeated neighbor counting
or local statistics, but only within bounded support rather than across the
full trajectory or subset space.

\item \textbf{Util-Stable.}
Util-Stable reflects whether the underlying evaluation functional admits an
explicit dataset-independent bound on worst-case marginal change. Coalition
methods score \Empty\ because utilities of the form
$U(S\cup\{z_j\})-U(S)$ can be arbitrarily large for rare or adversarial
coalitions, even when average behavior is benign. Smooth loss-based methods
(influence and trajectory estimators) score \Half\ since their stability can
sometimes be improved through gradient clipping or Lipschitz assumptions, but
no bound is inherent in the estimator. Locality-constrained utilities, such as
Threshold-$k$NN, score \Full\ because truncating contributions to a bounded
neighborhood yields explicit control over worst-case marginal effects.

\item \textbf{Sens-Driver.}
Sens-Driver identifies the dominant mechanism by which a single record can be
amplified into a large valuation score. Influence estimators are driven by
curvature inversion: even small per-example gradients can be magnified along
ill-conditioned Hessian directions (captured by $\|H^{-1}\|$). Coalition-based
methods are dominated by subset extrema: privacy is governed by
$\sup_S |\Delta_U|$, since rare coalitions can induce arbitrarily large marginal
utility changes. Trajectory methods are driven by accumulation across steps:
even moderate per-step effects can compound over many iterations. Surrogate
methods are driven by embedding norm products: sensitivity depends on
$\sup_z\|a(z)\|\cdot \sup_{z_j}\|s(z_j)\|$, which is rarely controlled. In
locality-based valuation, amplification is mediated through bounded
neighborhood size, which weakens global sensitivity drivers.

\item \textbf{Scale.}
Scale summarizes practical feasibility at modern model sizes. Exact influence
via iHVP is marked \Empty\ because computing $H^{-1}$ is prohibitive. Fisher
approximations, TracIn, and learned surrogates are marked \Full\ because they
admit efficient implementations that scale to large neural networks. This
column is included because privacy mechanisms are only meaningful if the
underlying valuation estimator is computationally deployable.

\end{squishitemize}

\medskip
\noindent\textbf{Privacy--sensitivity columns.}
The rightmost block isolates how each method behaves when viewed through the
lens of worst-case privacy.

\begin{squishitemize}

\item \textbf{Emp-Smooth.}
Emp-Smooth captures whether the estimator exhibits empirical attenuation of
extreme values through averaging, projection, surrogate modeling, or locality
constraints. Many scalable approximations score \Full\ here: Fisher surrogates
smooth curvature, projection suppresses small-eigenvalue directions, trajectory
methods average gradient interactions across steps, and learned surrogates
aggregate signal statistically. However, this smoothing is typically heuristic
and does not imply a certified DP bound.

\item \textbf{Formal-Bound.}
Formal-Bound indicates whether the estimator admits a principled,
dataset-independent worst-case sensitivity bound. Most influence, coalition,
trajectory, and surrogate methods score \Empty\ because their amplification
mechanisms (curvature gain, subset extrema, trajectory accumulation, or
embedding norms) remain uncontrolled in the worst case. Only explicitly
sensitivity-constrained designs---such as locality-truncated valuation---score
\Full, since bounding neighborhood scope yields direct worst-case control.

\item \textbf{Clip-Amenable.}
Clip-Amenable reflects whether bounded sensitivity can plausibly be enforced by
clipping, truncation, or norm constraints without destroying valuation
semantics. Many estimators are partially compatible (\Half): influence scores
can be clipped, gradients can be bounded, and surrogate embeddings can be
normalized. However, clipping is less meaningful for coalition methods where
extreme marginal contributions are inherent, and most methods lack a natural
calibration procedure. Locality-based methods score \Full\ because truncation
is intrinsic to their definition.

\end{squishitemize}

\paragraph{Family-level takeaways.}
Table~\ref{tab:sok_unified_privacy_block} emphasizes that different valuation
paradigms fail DP for fundamentally different reasons. Influence estimators are
limited by curvature amplification, coalition-based values by worst-case subset
effects, trajectory methods by compositional exposure across training steps,
and surrogate approaches by uncontrolled embedding sensitivity. While many
approximations yield substantial \emph{empirical} smoothing, this does not
translate into certified worst-case privacy guarantees. Locality-based and
explicitly sensitivity-constrained designs stand out as one of the few regimes
where structural restrictions yield principled control over single-record
influence.

Overall, the table motivates our central thesis: the key obstacle to private
data valuation is not merely computational approximation, but the lack of
worst-case sensitivity control over the dominant amplification drivers induced
by each estimator family.

%% file: sections/app1.tex
\input{sections/3.5_fixed}

%% file: sections/3.5_fixed.tex
\section{Locality-Based Approaches}
\label{sec:locality}

The approaches above estimate data value either through global sensitivity to
training dynamics (influence- and trajectory-based methods) or through marginal
contributions aggregated across subsets (Shapley-style methods). A
complementary perspective instead grounds valuation in the \emph{local geometry}
of representation space: examples are valuable insofar as they provide nearby
support for the model’s prediction at an evaluation point $z$.

From a privacy perspective, locality is appealing because it restricts the
scope of any single record’s influence. Contributions are confined to bounded
neighborhoods rather than amplified globally through curvature operators,
trajectory accumulation, or exponentially many subsets. This enables more
direct sensitivity control via truncation or clipping of local contributions.

\smallskip

\noindent\textbf{Approach 11: Threshold-$k$NN Shapley (local marginal contributions).}
Classical Shapley valuation is computationally infeasible at scale since it
requires marginal utilities over exponentially many subsets. For $k$-nearest
neighbors ($k$NN), however, \citet{jia2019efficient} derive a closed-form
Shapley computation. For validation $z=(x,y)$, the utility of a subset
$S\subseteq\train$ is the fraction of the $k$ nearest neighbors in $S$ that
share the correct label:
\[
U(S)
=
\frac{1}{k}\sum_{i=1}^{\min(k,|S|)} \mathbf{1}[y_i = y].
\]

While efficient, this utility has $O(1)$ sensitivity: adding or removing a
single close neighbor can abruptly change the vote, making the induced Shapley
values highly outlier-dependent and difficult to privatize~\cite{wang2023privacy}.

To enforce locality, \citet{wang2023privacy} propose \emph{Threshold-$k$NN
Shapley}, which replaces fixed-$k$ neighborhoods with a bounded radius. Let
$B_{x,\tau}=\{x' : d(x',x)\le \tau\}$ and define
$B_{x,\tau}(S)=\{(x_i,y_i)\in S : x_i\in B_{x,\tau}\}$. The thresholded utility
is
\[ 
U_\tau(S)
=
\begin{cases}
C, & |B_{x,\tau}(S)|=0,\\[0.3em]
\frac{1}{|B_{x,\tau}(S)|}\sum_{(x_i,y_i)\in B_{x,\tau}(S)} \mathbf{1}[y_i=y],
& |B_{x,\tau}(S)|>0,
\end{cases}
\]
where $C$ is a stable baseline (e.g., random-guess accuracy). Truncation
filters distant outliers and yields a more controlled local utility.

 key observation is that the resulting valuation can be expressed as a
function of low-dimensional neighborhood count statistics,
\[ 
v_j^{(\tau)}(z)=f\!\left(C_z(\train\setminus\{z_j\})\right),
\]
enabling differential privacy by privatizing these local counts and invoking
post-processing closure. More broadly, locality-based methods illustrate how
restricting attribution to bounded neighborhoods can provide more explicit
sensitivity control than global Shapley or influence estimators.